\documentclass[
 reprint,
 amsmath,amssymb,
 aps,
]{revtex4-2}

\usepackage{graphicx}
\usepackage{epstopdf}
\usepackage{ulem}
\usepackage{dcolumn}
\usepackage{bm}

\usepackage{psfrag}
\usepackage{natbib}
\usepackage{tikz}
\usepackage{upgreek}
\usepackage{multirow}
\usepackage{subfigure}
\usepackage{color}
\usepackage{amsmath}
\usepackage{amssymb}
\usepackage{afterpage}
\usepackage{array}
\usepackage{floatrow}
\usepackage{comment}
\usepackage{enumitem}
\usepackage{tikz}
\usetikzlibrary{shapes,arrows,positioning,calc}

\newfloatcommand{capbtabbox}{table}[][\FBwidth]
\usepackage[colorlinks=true, linkcolor=blue, urlcolor=blue, citecolor=blue]{hyperref}

\definecolor{blue2}{rgb}{0, 0.4, 0.7}
\definecolor{green2}{rgb}{0, 0.6, 0}
\def\drwln#1#2{\raise 2.5pt\vbox{\hrule width #1pt height #2pt}}
\definecolor{Gr}{rgb}{0, 0.5, 0.1} 

\begin{document}

\preprint{APS/123-QED}

\tikzset{every picture/.style={line width=1pt}}
\tikzstyle{block} = [draw, fill=white, rectangle, 
    minimum height=3em, minimum width=6em]
\tikzstyle{sum} = [draw, fill=white, circle, node distance=1cm]
\tikzstyle{input} = [coordinate]
\tikzstyle{output} = [coordinate]
\tikzstyle{pinstyle} = [pin edge={to-,thin,black}]

\title{Streak generation in viscosity-stratified wall-bounded flows}

\author{Anagha Madhusudanan}
 \email{anagham@iisc.ac.in, anaghamadhu91@gmail.com}
\affiliation{%
 Department of Aerospace Engineering, Indian Institute of Science, Bengaluru 560012, India 
}%
\author{Simon J. Illingworth}
\affiliation{
  Mechanical Engineering, University of Melbourne, VIC 3010, Australia
}%
\author{Rama Govindarajan}
\affiliation{%
  International Centre for Theoretical Sciences,  Tata Institute of Fundamental Research, Bengaluru 560089, India
}%

\date{\today}

\begin{abstract}
Streamwise elongated flow structures, or streaks, dominate wall-bounded flows. 
We show that the renowned lift-up mechanism, that generates streaks, is significantly altered by viscosity stratification. 
We additionally identify a novel pathway for streak generation: the viscosity-stratified lift-up mechanism. 
The competition and cooperation of the streaks generated by these mechanisms provides an explanation for trends in viscosity-stratified Poiseuille flows, and remarkably, the opposing trends in Couette flows as well. 
Our theoretical observations are substantiated by numerical experiments. 
\end{abstract}

\maketitle

Streaks, i.e., elongated flow structures, are characteristic of wall-bounded flows that are turbulent or transitioning to turbulence \citep{kline1967structure,smits2011high}. 
These streaks play crucial roles in the statistics and dynamics of wall-bounded flows (see review \citep{smits2011high}), and controlling these flows often involves controlling the streaks. 
Understanding these streaks is therefore crucial for a range of situations: from applications like drag reduction in wall-bounded flows \citep[e.g.][]{kuhnen2018destabilizing, marusic2021energy} to atmospheric flows \citep[e.g.][]{hutchins2012towards} and hurricanes \citep[e.g.][]{momen2021scrambling}. 

In constant viscosity flows, the generation of these streaks is well characterized by a classical mechanism -- the lift-up mechanism -- where vortices in the flow extract energy from the mean shear to generate velocity streaks \citep{landahl1980note, ellingsen1975stability}. 
However, in many practically relevant flows, viscosity stratification, i.e., a variation in the viscosity of the fluid with temperature or scalar concentration, is important: flows such as the flow in the Earth's core \citep[e.g.][]{steffen2005glacial}, crude oil flow in pipelines \citep[e.g.][]{beal1946viscosity} and polymer-laden flows \citep[e.g.][]{kalashnikov1994shear}. 
Fully compressible flows, like flows over reentry vehicles \citep[e.g.][]{wilke1950viscosity}, are also impacted by viscosity stratification. 
Importantly, in viscosity stratified flows, we find that the classical mechanism does not sufficiently explain the streaks. 
Here we present an entirely new mechanism and an altered lift-up mechanism that are now in operation. 
The interaction of these mechanisms leads to a fundamental revision of our understanding of streak generation in these flows, and also produces some interesting effects, including some counterintuitive ones. 

To illustrate these mechanisms, let us consider incompressible channel flows, i.e.\ flows between two parallel walls. 
The bottom cold wall is at a temperature $\Theta_C^*$ and the top hot wall is at temperature $\Theta_H^*$, with temperature difference $\Delta\Theta^* = \Theta_H^* - \Theta_C^*$ 
(here $(\cdot)^*$ denotes dimensional quantities). 
We consider the Poiseuille flow, driven by a pressure gradient, and the Couette flow, driven by the velocity of the wall.  
Here we consider the case where the viscosity of the fluid $\mu(\theta)$ varies with temperature $\theta$. 
Note, the mechanisms described here will also impact flows where the viscosity is a function of scalar concentration. 
Mathematical models have shown that viscosity stratification alters wall-bounded flows \citep[e.g.][]{govindarajan2001retardation, chikkadi2005preventing, govindarajan2014instabilities, thakur2021early, rinaldi2018edge}.
Direct numerical simulations (DNSs) have also demonstrated that viscosity stratification modifies turbulence \citep[e.g.][]{lee2013effect, zonta2012modulation}.  
For instance, while intuitively, higher viscosity would be expected to suppress turbulence, DNS studies have shown an enhancement of turbulence in the more-viscous regions of some flows, and a relative suppression in the less-viscous regions \citep[e.g.][]{zonta2012modulation, lee2013effect}. 
The mechanisms presented here explain this surprising trend. 

The streamwise, spanwise and wall-normal directions of the channel are $x$, $y$ and $z$ and the corresponding velocity components are $u$, $v$ and $w$. 
The mean velocity $(U(z),0,0)$ and the mean temperature $\Theta(z)$ are obtained by averaging in $x$, $y$ and time. 
The Reynolds number is $Re\equiv \rho^*_{r}U_b^*h^*/\mu_{r}^*$, and the Prandtl number is $Pr\equiv \mu_{r}^*c_{p,r}^*/\lambda_{r}^*$. 
Here $U_b$ is the bulk velocity, $h$ is the channel half height, $\rho$ is density, $c_{p}$ is the specific heat at constant pressure and $\lambda$ is the thermal conductivity. 

Streaks can be analyzed using linearized Navier--Stokes equation-based models \citep{jovanovic2005componentwise, illingworth2018estimating}. 
To obtain the model, we linearize the non-dimensional Navier--Stokes equations around the mean flow. 
For a majority of the discussion, we linearize around the laminar mean flow, except in figure \ref{fig:AmpTurb_R1R2} where we show that the conclusions drawn in this study are also more generally valid when the linearization is done around the turbulent mean flow. 
We Fourier transform the state variables in $x$, $y$ and time. 
Here $\widehat{\cdot}$ represents the Fourier transform.  
The streamwise and spanwise wavenumbers are ($k_x,k_y$) and the temporal frequency is $\omega$ ($\omega$ is multiplied by $Re$ for ease of notation).  
Since streaks are elongated structures, we consider the response from streamwise-constant ($k_x=0$) structures. 
This streamwise-invariant model then becomes: 
\begin{equation}
\begin{split}
\frac{1}{Re} \bm{L}_{OS} \widehat{w} &= 
-ik_y\frac{\partial }{\partial z}\widehat{f}_y - k^2\widehat{f}_z, 
\\
\frac{1}{Re} \bm{L}_{SQ} \widehat{u} &= 
-   U' \widehat{w}
+\frac{1}{Re} \bm{M}_{u \theta} \widehat{\theta} 
+ \widehat{f}_x,  
 \\
\frac{1}{RePr} \bm{L}_\Theta \widehat{\theta} &= 
- \Theta' \widehat{w}
+ \widehat{f}_\theta. \\
\end{split}
\label{eqn:OrrSommerfeldSquire}
\end{equation}
Here $(f_x,f_y,f_z)$ and $f_\theta$ represents any external forcing, and, for turbulent flows, the nonlinear terms of the equations. 
The boundary conditions imposed are $\widehat{w}(\pm1)=\partial\widehat{w}/\partial z(\pm1)= (ik_y\widehat{u}-ik_x\widehat{v})(\pm1)=\widehat{\theta}(\pm1)= 0$. 
The linear operators in \eqref{eqn:OrrSommerfeldSquire} are: 
\begin{equation}
\begin{split}
\bm{L}_{OS} &= i\omega\Delta - \overline{\mu} \Delta^2 
-2\overline{\mu}' \mathcal{D} \Delta
+ \overline{\mu}'' \left( \Delta 
-2 \mathcal{D}^2 \right) , \\
\bm{L}_{SQ} &= i\omega -  \overline{\mu} \Delta 
- \overline{\mu}' \mathcal{D} , \\
\bm{M}_{u \theta} &= \overline{\varepsilon} \mathcal{D} 
+ \overline{\varepsilon}'  \quad \mbox{ and } \quad 
\bm{L}_\Theta= i\omega Pr - \Delta,
\end{split}
\label{eqn:OrrSommerfeldSquire_LOS_LSQ}
\end{equation}
where $'$ and $\mathcal{D}$ represent derivatives in $z$, $\Delta=\nabla^2$ and $\overline{\varepsilon} = U'(d \overline{\mu}/d \Theta)$.  
Figure \ref{fig:BlockDiag_StreamAmp_ViscStrat} shows \eqref{eqn:OrrSommerfeldSquire} as a block diagram. 

\begin{figure}
\resizebox{\textwidth}{!}{
\begin{tikzpicture}[auto, node distance=1.5cm,>=latex']     
	\node [input, name=input] {};
    \node [sum, right of=input] (sum) {};
    \node [block, right of=sum] (controller) {\LARGE $\bm{L}_{OS}^{-1}$};
    \node [block, left of=sum, minimum height=2em, minimum width=3em] (vec2) {\LARGE $-k^2$};
    \node [block, above = 0.25cm of vec2, minimum height=2em, minimum width=3em] (vec1) {\LARGE $-ik\mathcal{D}$};
    \node [block, right = 0.5cm of controller, minimum height=2em, minimum width=3em] (R1) {\LARGE $Re$};
    \node [block, right = 1.1cm of R1, minimum height=2em, minimum width=3em] (controller2_1) {\LARGE $-U'$};
    \node [sum, right = 0.5cm of controller2_1] (sum2) {};
    \node [block, right=1cm of sum2] (controller2) {\LARGE $\bm{L}_{SQ}^{-1}$};
    \node [block, right = 0.5cm of controller2, minimum height=2em, minimum width=3em] (R2) {\LARGE $Re$}; 
    
    \node [block, below = 1.6cm of controller2_1, minimum height=2em, minimum width=3em] (controllerT_1) {\LARGE $-\Theta'$};
    \node [sum, right = 0.5cm of controllerT_1] (sum3) {};
    \node [block, right=1cm of sum3] (controllerT) {\LARGE $Pr\bm{L}_\Theta^{-1}$};
    
    \node [block, above = 0.3cm of controller2, minimum height=2em, minimum width=3em, draw=black!25!white] (vec4) {\textcolor{black!25!white}{\LARGE $i\mathcal{D}/k$}};
    \node [block, right = 0.5cm of controllerT, minimum height=2em, minimum width=3em] (R3) {\LARGE $Re$}; 
    \node [block, above = 0.15cm of controllerT, minimum width=3em] (MuT) {\LARGE $\bm{M}_{u\theta}$}; 
    
    \node [output, right of=R2, node distance=2cm] (output) {};
    \node [output, right of=R3, node distance=2cm] (outputT) {};
    \node [output, right of=vec4, node distance=4.03cm] (output2) {};
    \node [output, above = 1cm of output2] (output3) {};
    \node [input, left = 1cm of vec2] (input1) {};
    \node [input, left = 0.7cm of vec1] (input2) {};
    \node [input, above = 1.5cm of input2] (input3) {};
    \node [input, below = 1.5cm of input1] (inputT) {};
    \node [input, below = 1cm of input1] (input4) {};
    
    \draw [->] (input1) --node[left = 0.5cm of input1]{\textcolor{red!90!black}{\huge $\widehat{f}_z$}} (vec2);
    \draw [->] (input2) --node[name=y, left = 0.5cm of input2]{\textcolor{red!90!black}{\huge $\widehat{f}_y$}} (vec1);
    \draw [->] (input3) node[left = 0.2cm of input3]{\textcolor{red!90!black}{\huge $\widehat{f}_x$}} -| (sum2);
    \draw [->] (inputT) node[left]{\textcolor{red!90!black}{\huge $\widehat{f}_\theta$}} -| (sum3);    
    \draw [->] (vec1) -| (sum);
    \draw [->] (vec2) -- (sum);
    \draw [->] (sum) -- (controller);
    \draw [->] (controller) -- (R1);
    \draw [->] (R1) --node[above]{\huge $\widehat{w}$} (controller2_1);
    \draw [->] (controller2_1) -- (sum2);
    \draw [->] (sum2) -- (controller2);
    \draw [->] (controller2) -- (R2);
    \draw [->] (R2) --node[pos=1]{\huge $\widehat{u}$} (output);
    \draw [<-,black!25!white] (vec4) -| ($(1cm,0)+(R1)$);
    \draw [->,black!25!white] ($(1cm,0)+(R1)$) |-node[pos=1]{\LARGE $\widehat{w}$} (output3);
    \draw [->,black!25!white] (vec4) --node[pos=1]{\huge $\widehat{v}$} (output2);
    \draw [->] (R3) --node[pos=1]{\huge $\widehat{\theta}$} (outputT);
    \draw [<-] (MuT) -| ($(-0.9cm,0cm)+(R3)$);  
    \draw [->] (MuT) -| (sum2);    
    \draw [<-] (controllerT_1) -| ($(1cm,0)+(R1)$);
    \draw [->] (controllerT_1) -- (sum3);
    \draw [->] (sum3) -- (controllerT);
    \draw [->] (controllerT) -- (R3);
    
\end{tikzpicture}
}
\caption{Block diagram representing equations \eqref{eqn:OrrSommerfeldSquire}.} 
\label{fig:BlockDiag_StreamAmp_ViscStrat}
\end{figure} 
To understand streak generation, we study how $\widehat{u}$ is generated by the inputs $\widehat{f}_x$, $\widehat{f}_y$, $\widehat{f}_z$ and $\widehat{f}_\theta$. 
Tracing out the different pathways in figure \ref{fig:BlockDiag_StreamAmp_ViscStrat}, $\widehat{u}$ is obtained as:
\begin{equation}
    \begin{split}
        \widehat{u} = 
        &Re^2\bm{R}_1(z;k_y,\omega) 
        \begin{bmatrix}  f_y \\ f_z    \end{bmatrix}  +
        Re^2\bm{R}_2(z;k_y,\omega) 
        \begin{bmatrix}  f_y \\ f_z    \end{bmatrix}  +\\
        &Re \bm{R}_3(z;k_y,\omega) 
        \begin{bmatrix}  f_x    \end{bmatrix}  +
        Re \bm{R}_4(z;k_y,\omega) 
        \begin{bmatrix}  f_\theta    \end{bmatrix},
    \end{split}
    \label{eqn:Ruf}
\end{equation}
where the linear operators are given by
\begin{equation}
    \begin{split}
        \bm{R}_1 &= -\bm{L}_{SQ}^{-1} U' \bm{L}_{OS}^{-1} [-ik\mathcal{D}, -k^2], \\
        \bm{R}_2 &= -Pr \bm{L}_{SQ}^{-1} \bm{M}_{u\theta} \bm{L}_{\Theta}^{-1} \Theta' \bm{L}_{OS}^{-1} [-ik\mathcal{D}, -k^2], \\
        \bm{R}_3 &= -\bm{L}_{SQ}^{-1}, \quad \mbox{and} \quad
        \bm{R}_4 = -Pr\bm{L}_{SQ}^{-1} \bm{M}_{u\theta} \bm{L}_{\Theta}^{-1}.  \\
    \end{split}
    \label{eqn:R1R2}
\end{equation}
From \eqref{eqn:Ruf}, the contributions from $\bm{R}_3 \widehat{f}_x$ and $\bm{R}_4 \widehat{f}_\theta$ scale with $Re$, whereas the contributions from $(\widehat{f}_y,\widehat{f}_z)$, via $\bm{R}_1$ and $\bm{R}_2$, scale with $Re^2$.  
At moderate and high Reynolds numbers, the contributions of $\bm{R}_3$ and $\bm{R}_4$ can therefore be ignored.  
$\bm{R}_1$ corresponds to streak formation by the classical lift-up mechanism, albeit modulated by viscosity stratification. 
$\bm{R}_2$ corresponds to a completely new viscosity-stratified lift-up mechanism driven by the temperature gradient. 
Unlike in the case of constant viscosity flows, in viscosity stratified flows, the streaks generated by $\bm{R}_1$ and $\bm{R}_2$ interact. 
Crucially, the contribution of $\bm{R}_2$ to streak amplification increases with $Pr$, and therefore, at high enough $Pr$, it is likely that this novel viscosity-stratified lift-up becomes the dominant mechanism. 

To illustrate how $\bm{R}_1$ and $\bm{R}_2$ operate, we will now consider a specific function for viscosity $\mu(\theta)$ as
\begin{equation}
\mu(\theta) = \exp{\left( A \left(\dfrac{B^*}{(\Theta_{r}^*-C^*)} \right) \dfrac{(1-\theta)}{(\theta-C^*/\Theta_{r}^*)} \right)}, 
\label{eqn:visc}
\end{equation}
where $A=\log_e(10)$, $B^*=247.8K$, $C^*=140K$ and the reference temperature $\Theta_{r}^*=(\Theta_H^*+\Theta_C^*)/2=323K$ \citep{zonta2012modulation}. 
The viscosity is here non-dimensionalized by $\mu_{r}$, where a subscript $(\cdot)_r$ denotes reference quantities defined at $\Theta_r^*$. 
This exponential law is an empirical fit frequently employed for liquids like water \citep{zonta2012modulation, lee2013effect}.  
The streak amplification mechanisms discussed here remain largely unaffected if the viscosity instead varies according to Sutherland's law, commonly used for fluids like air. 

$\bm{R}_1$, $\bm{R}_2$ and $(\bm{R}_1+\bm{R}_2)$ are linear operators that take an input forcing $\widehat{\bm{f}}=(\widehat{f}_y,\widehat{f}_z)$ and give a response $\widehat{u} = (\bm{R}_1+\bm{R}_2)\widehat{\bm{f}}$. 
To understand these operators, we can ask what the most sensitive forcing is, i.e.\ what is the $\widehat{\bm{f}}$ that gives the $\widehat{u}$ with maximum energy? 
For this we use singular value decomposition (SVD), which gives
$(\bm{R}_1 + \bm{R}_2) = \sum_{i=0}^{N} \psi_i \sigma_i \bm{\phi}_i$,
with $\sigma_{i} \geq \sigma_{i+1}$ and unit norm vectors $\bm{\phi}_i$ and $\psi_i$. 
In effect, for $i=1$, we have forcing $\widehat{\bm{f}}=\bm{\phi}_1$ that gives the maximum response $\widehat{u} = \sigma_1\psi_1 = (\bm{R}_1+\bm{R}_2) \bm{\phi}_1$. 

Given these definitions at a temporal frequency $\omega$, we can now ask what the maximum possible response is across frequencies.
For this, we use the infinity norm:
\begin{equation}
\begin{split}
||(\bm{R}_1+\bm{R}_2)||_\infty(k_y) = \max_\omega \mbox{ } \sigma_1(k_y,\omega). 
\end{split}
\end{equation}
Let the $\omega$ for the maximum response be $\omega_{m}$,  
the corresponding forcing $\bm{\phi}_m$ and the response $\psi_m$. 
Therefore, $\bm{\phi}_m$ is the optimum forcing $(\widehat{f}_y,\widehat{f}_z)$ that, across all frequencies, generates the maximum response $\widehat{u}=\sigma_m\psi_{m}$. 

Now, we compute the separate contributions of $\bm{R}_1$ and $\bm{R}_2$ at $\omega_m$ as $\psi_{R_1} = \bm{R}_1 \bm{\phi}_m$ and $\psi_{R_2} = \bm{R}_2 \bm{\phi}_m$. 
We also consider the distinct contributions from the more-viscous and less-viscous channel halves. 
The energy of these responses numerically obtained on $N=101$ wall-normal Chebyshev grid points is considered. 

\begin{figure}
\subfigure{
\centering
\includegraphics[width=0.9\textwidth]{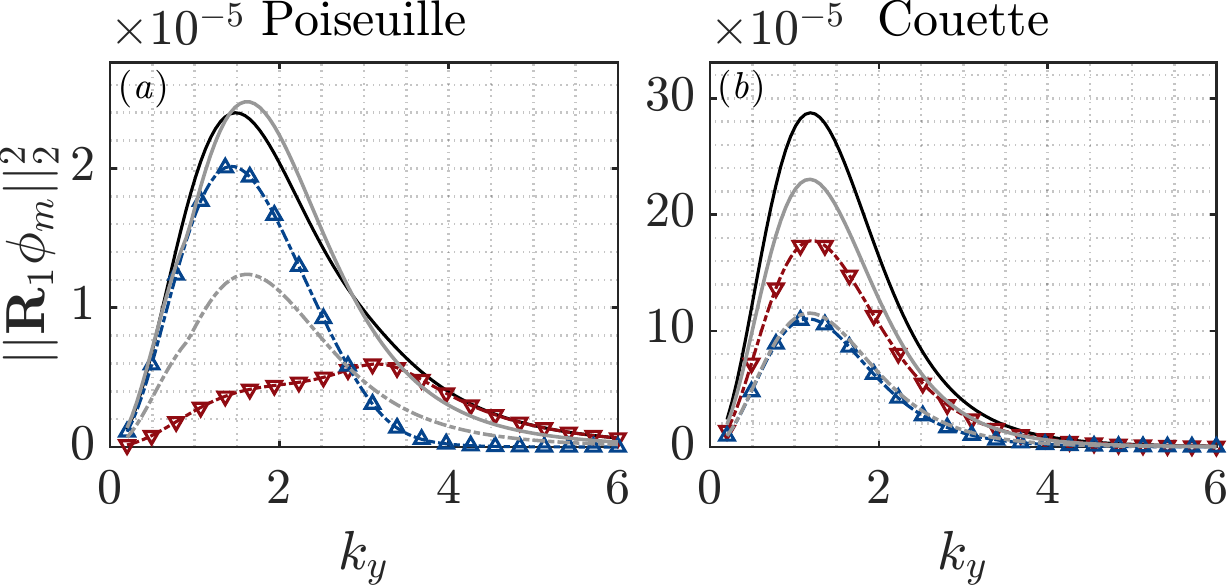}
\label{fig:Amp_R1_chan}
}%
\subfigure{
\centering
\label{fig:Amp_R1_couette}
}
\caption{
For (\textit{a}) Poiseuille and (\textit{b}) Couette flows, at $Pr=3$, $Re=500$ and $\Delta\Theta^* = 40K$, the contribution of the classic lift-up mechanism $||\bm{R}_1\bm{\phi}_m||_2^2$ is shown as a function of $k_y$. 
The unstratified flow at $Re=500$ (gray lines) is also shown. 
The responses across the full channel (solid lines) and across channel halves (dashed lines) are shown. 
The more-viscous ($\triangle$ markers) and less-viscous ($\nabla$ markers) channel halves are compared against the unstratified channel half (dashed gray).  
}
\label{fig:Amp_R1}
\end{figure}

\underline{Route R1: Classical lift-up mechanism} --
We now examine how the well-known lift-up mechanism in unstratified flows is modified by viscosity stratification.  
In lift-up, the wall-normal velocity ($\widehat{w}$) leverages the mean shear ($U'$) to generate streamwise velocity streaks \citep[e.g.][]{landahl1980note,ellingsen1975stability}. 
From figure \ref{fig:BlockDiag_StreamAmp_ViscStrat} and \eqref{eqn:R1R2}, this corresponds to the transfer function $\bm{R}_1$. 
Viscosity stratification modifies $\bm{R}_1$ in two ways: through (i) additional terms in the linear operators due to viscosity and its gradients, and (ii) the modification of the mean shear. 
Figure \ref{fig:Amp_R1} shows the contribution of $\bm{R}_1$ as a function of the spanwise wavenumber  $k_y$. 

When considering the channel halves separately, for Poiseuille flows in figure \ref{fig:Amp_R1}\subref{fig:Amp_R1_chan}, viscosity stratification enhances the lift-up mechanism in the more-viscous channel half and diminishes it in the less-viscous half. 
Counterintuitively, therefore, higher viscosity is enhancing streaks rather than suppressing them. 
This observation is consistent with the DNS of turbulent viscosity-stratified Poiseuille flows of \citet{zonta2012modulation}. 
Remarkably, this trend is absent for the Couette flow in figure \ref{fig:Amp_R1}\subref{fig:Amp_R1_couette}, where streaks are more energetic in the less-viscous half. 
Additionally, for the Poiseuille flow, there is a mismatch between the optimal wavenumbers in the two channel halves, which is absent in the Couette flow. 

\begin{figure}
\captionsetup[subfigure]{labelformat=empty,skip=-10pt}
\subfigure{
\centering
\centering
\includegraphics[width=0.9\textwidth]{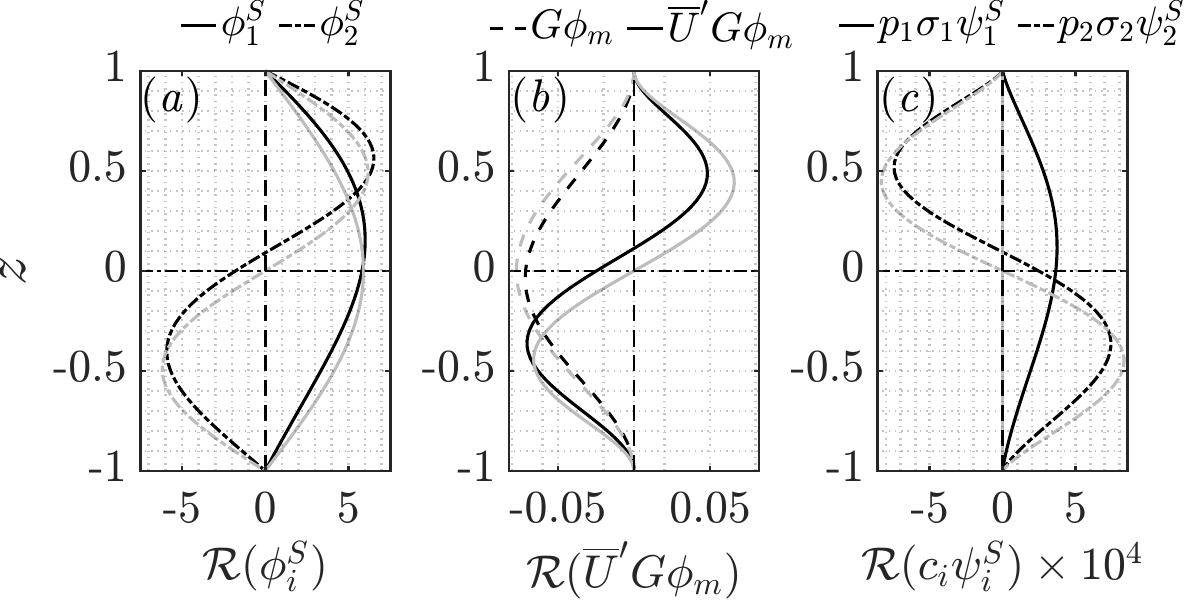}
\label{fig:R1_modes_Gphi}
}%
\subfigure{
\centering
\label{fig:R1_modes_Sphi}
}%
\subfigure{
\centering
\label{fig:R1_modes_Spsi}
}
\caption{
For $(k_x,k_y) = (0,\pi/2)$ in viscosity-stratified Poiseuille flows at $Re=500$, $Pr=3$ and $\Delta\Theta^* = 40K$ (black lines) and the unstratified flow at $Re=500$ (grey lines), the real part of (\textit{a}) $\bm{\phi}_1^S$ and $\bm{\phi}_2^S$, (\textit{b}) $\bm{G} \bm{\phi}_m$ and $U'\bm{G} \bm{\phi}_m$, (\textit{c}) $(p_1\sigma_1^S\psi_1^S)$ and $(p_2\sigma_2^S\psi_2^S)$ are shown. 
Here $\mathcal{R}(\cdot)$ represents real part. 
(Note, $\mathcal{R}(p_1)\leq0$ and $\mathcal{R}(p_2)>0$.)
}
\label{fig:R1_modes}
\end{figure}
The question that follows is what causes this contrasting trend in Poiseuille and Couette flows?
To highlight the effect of shear in $\bm{R}_1$, let us write $\bm{R}_1  =  - \bm{S} \cdot U' \cdot \bm{G}$. 
The operators $\bm{S}$ and $\bm{G}$ follow from \eqref{eqn:R1R2} and are identical for Poiseuille and Couette flows. 
Using SVD $\bm{S}=(\sum_{i=1}^N \psi^S_i \sigma^S_i \phi^S_i$),  
and the response to $\bm{R}_1$ becomes:
\begin{equation}
    \psi_{R_1} = \bm{R}_1 \bm{\phi}_m = \sum_{i=1}^N \psi^S_i \mbox{ } \sigma^S_i \mbox{ }\underbrace{(\phi^S_i  \cdot U' \bm{G}\bm{\phi}_m)}_{p_i}.
    \label{eqn:R1_rewrite}
\end{equation}
The contribution of a mode $i$ in the summation is determined by $p_i$ that is the projection of $U'\bm{G}\bm{\phi}_m$ onto the $\phi_i^S$. 
Since $\sigma^S_1 > \sigma^S_i, i=2,3, ....N$, the lift-up represented by $\bm{R}_1$ is most efficient when $p_1$ is the maximum. 

Let us first consider unstratified flows as in \citet{illingworth2020streamwise}. 
The wall-normal profile of $\phi_1^S$ is symmetric across $z=0$, and $\phi_2^S$ is antisymmetric (figures \ref{fig:R1_modes}(\textit{a})). 
$\bm{G}\bm{\phi}_m$ is symmetric (figures \ref{fig:R1_modes}(\textit{b})). 
These profiles are identical for Poiseuille and Couette flows. 
In Couette flows, $U'$ is symmetric, making $U'\bm{G}\bm{\phi}_m$ symmetric. 
The symmetric $U'\bm{G}\bm{\phi}_m$ projects well onto the symmetric $\phi_1^S$, thereby producing a large $p_1$. 
Lift-up is therefore efficient in Couette flows \citep{illingworth2020streamwise}. 
In Poiseuille flows however, $U'$ is antisymmetric, and therefore so is $U'\bm{G}\bm{\phi}_m$ (figures \ref{fig:R1_modes}(\textit{b})). 
The antisymmetric $U'\bm{G}\bm{\phi}_m$ does not project onto the symmetric $\phi_1^S$ and instead projects onto the antisymmetric $\phi_2^S$. 
The projection $p_1=0$, and the contribution of the first mode $p_1\sigma_1\psi_1=0$ (figures \ref{fig:R1_modes}(\textit{c})). 
Lift-up in Poiseuille flows is therefore inefficient compared to Couette flows \citep{illingworth2020streamwise}. 

Now let us follow the same arguments for viscosity-stratified flows. 
Viscosity stratification introduces asymmetry. 
Therefore, in Poiseuille flows, $U'\bm{G}\bm{\phi}_m$ is nearly antisymmetric, but not perfectly. 
The projection of $U'\bm{G}\bm{\phi}_m$ onto the symmetric $\phi_1^S$, although small, is now not zero, i.e.\ $p_1 >0$. 
Since $\sigma^S_1 \gg \sigma^S_2$, this small $p_1$ creates a significant contribution from the first mode $p_1\sigma_1\psi_1$ (figures \ref{fig:R1_modes}(\textit{c})). 
This now creates the scenario where the nearly-symmetric $p_1\sigma_1\psi_1$ and nearly anti-symmetric $p_2\sigma_2\psi_2$ interact, as illustrated in figure \ref{fig:R1_modes}(\textit{c}). 
We see that the two modes cooperate (have same sign) in the more-viscous bottom half, thereby enhancing streaks, and the modes compete (have opposite signs) in the less-viscous top half, thereby attenuating streaks. 
Therefore, the interaction of the symmetric and antisymmetric modes of $\bm{R}_1$ causes the enhancement of classical lift-up in the more-viscous region of Poiseuille flows.
Since classical lift-up is already optimal for unstratified Couette flows, this enhancement of lift-up due to viscosity-stratification-induced asymmetry is not significant in these flows. 

This comparison of Poiseuille and Couette flows therefore demonstrates that the classical lift-up mechanism is significantly modified by viscosity stratification. 
More generally, any asymmetry in the flow can strongly influence this key mechanism in shear flows. 
The observations also illustrate how the impact of this modification of classical lift-up varies across flow regimes. 

\underline{Route R2: viscosity-stratified lift-up mechanism} - 
Viscosity stratification also introduces an entirely new pathway for streak generation, separate from the classical lift-up mechanism.  
In this viscosity-stratified lift-up mechanism, the wall-normal velocity $(\widehat{w})$ leverages the mean temperature gradient $(\Theta')$ to generate a temperature streak $(\widehat{\theta})$, or equivalently, a viscosity streak. 
This in turn generates a streamwise velocity streak. 
From figure \ref{fig:BlockDiag_StreamAmp_ViscStrat} and \eqref{eqn:R1R2}, this corresponds to the transfer function $\bm{R}_2$. 
From \eqref{eqn:R1R2}, we note that, unlike $\bm{R}_1$, the contribution of $\bm{R}_2$ increases with $Pr$ and $\Delta \Theta^*$ ($\Theta'$ increases with $\Delta \Theta^*$). 

\begin{figure}
\captionsetup[subfigure]{labelformat=empty,skip=-15pt}
\subfigure{
\centering
\includegraphics[width=0.585\textwidth]{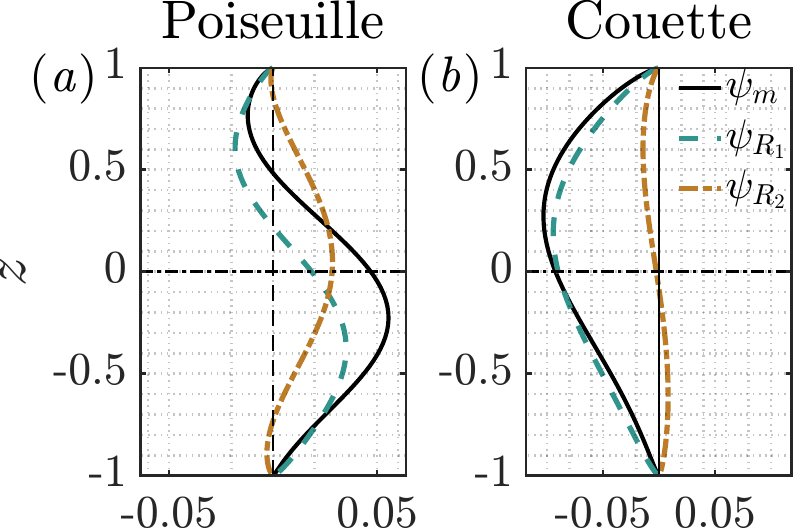} 
\label{fig:R1R2_chan}
}%
\subfigure{
\centering
\label{fig:R1R2_couette}
}
\caption{
The $\widehat{u}$ from responses $\psi_m$, $\psi_{R_1}$ and $\psi_{R_2}$ for the mode $(k_x,k_y,\omega) = (0,\pi/2,0)$ in a stratified channel flow at $Re=500$, $Pr=3$ and $\Delta\Theta^* = 40K$. 
Profiles from (\textit{a}) Poiseuille flow and (\textit{b}) Couette flow. }
\label{fig:R1R2}
\end{figure}

\begin{figure}
\captionsetup[subfigure]{labelformat=empty,skip=0pt}
\subfigure{
\centering
\centering
\includegraphics[width=0.9\textwidth]{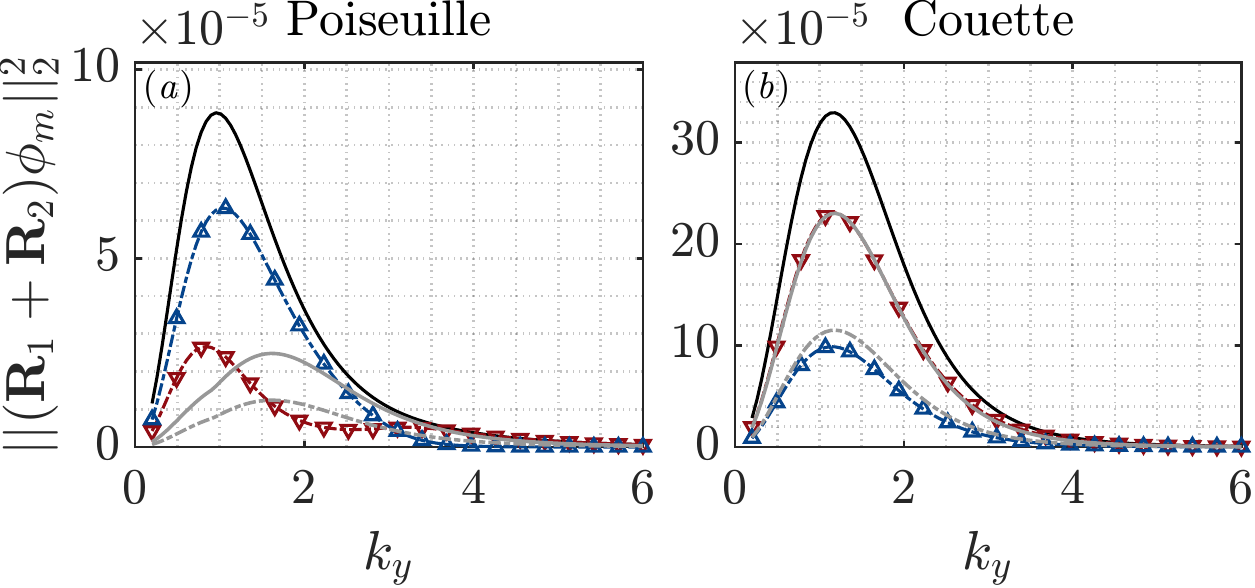}
\label{fig:Amp_R1R2_chan}
}%
\subfigure{
\centering
\label{fig:Amp_R1R2_couette}
}
\caption{
For (\textit{a}) Poiseuille and (\textit{b}) Couette flows, at $Pr=3$, $Re=500$ and $\Delta\Theta^* = 40K$, $||(\bm{R}_1+\bm{R}_2)\bm{\phi}_m||_2^2$ is shown with respect to $k_y$. 
The unstratified flow at $Re=500$ is also shown. 
The colors and markers represent the same as in figure \ref{fig:Amp_R1}. 
}
\label{fig:Amp_R1R2}
\end{figure}

This prompts us to ask how the separate contributions from R1 and R2 interact to give the full response. 
Figure \ref{fig:R1R2} shows the $\widehat{u}$ from R1, $\psi_{R_1}=\bm{R}_1\bm{\phi}_m$, and the $\widehat{u}$ from R2, $\psi_{R_2}=\bm{R}_2\bm{\phi}_m$, alongside the full response $\psi_m=(\psi_{R_1}+\psi_{R_2})=(\bm{R}_1+\bm{R}_2)\bm{\phi}_m$. 
Crucially, for the Poiseuille flow in figure \ref{fig:R1R2}\subref{fig:R1R2_chan}, the streaks from R1 and R2 cooperate (have same sign) at the more-viscous bottom channel half, thereby enhancing streaks, and they compete (have opposite signs) at the less-viscous top half, hence attenuating streaks.  
This interaction can cause streaks to be more energetic than in the unstratified case, with the major contribution to this energy coming from the more-viscous half of the channel, as seen in the trends of $||(\bm{R}_1+\bm{R}_2)||_\infty^2$ in figure \ref{fig:Amp_R1R2}\subref{fig:Amp_R1R2_chan}. 
The opposite trend is true for the Couette flow in figure \ref{fig:R1R2}\subref{fig:R1R2_couette}, with streaks from R1 and R2 cooperating at the less-viscous region of the channel, and therefore the less-viscous half showing higher energy in figure \ref{fig:Amp_R1R2}\subref{fig:Amp_R1R2_couette}. 
Additionally, since R1 is more efficient in Couette flows, in figure \ref{fig:R1R2}\subref{fig:R1R2_couette}, at a given $Pr$ and $\Delta \Theta^*$, the contribution of R1 is higher than for the Poiseuille flow. 
The competition and cooperation of R1 and R2 can therefore produce different, sometimes counterintuitive, trends in viscosity-stratified shear flows.  

From the model-based observations, we conclude that the asymmetry induced by viscosity stratification alters the classical lift-up mechanism for streak generation in shear flows. 
Additionally, an entirely new mechanism for streak generation, a viscosity stratified lift-up, gains dominance over the classical mechanism with increasing stratification. 
Illustrating the significance of this new and modified mechanisms, we found that the interaction of the streaks from the viscosity stratified lift-up with the altered classical lift-up produces contrasting trends in Poiseuille and Couette flows. 
In Poiseuille flows, the model predicts higher streak energy in the more-viscous regions; this trend is absent in Couette flows. 

\begin{figure}
\captionsetup[subfigure]{labelformat=empty,skip=0pt}
\subfigure{
\centering
\centering
\includegraphics[width=0.9\textwidth]{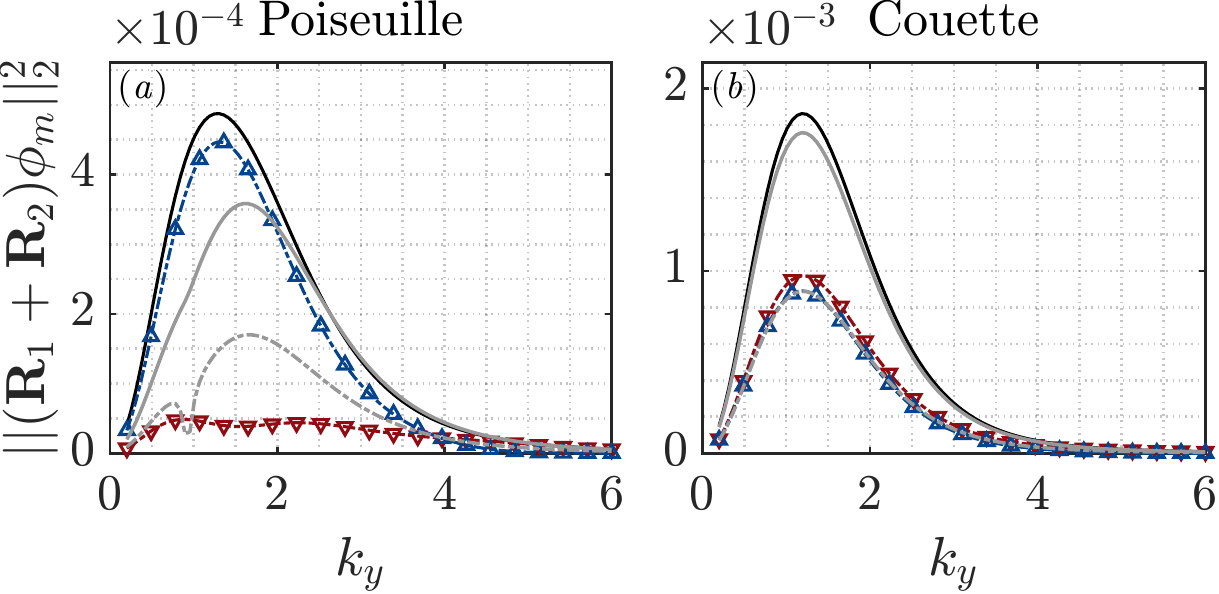}
\label{fig:AmpTurb_R1R2_chan}
}%
\subfigure{
\centering
\label{fig:AmpTurb_R1R2_couette}
}
\caption{
The same figure as \ref{fig:Amp_R1R2} for the linear model built using the turbulent mean profiles for (\textit{a}) a $Re_\tau=186$ Poiseuille and (\textit{b}) a $Re_\tau=100$ Couette flows, at $Pr=1$ and $\Delta\Theta^* = 40K$.  }
\label{fig:AmpTurb_R1R2}
\end{figure}

The final question we address here is, can this particular prediction from the model, i.e.\ the contrasting trends between Poiseuille and Couette flows, be demonstrated numerically? 
For this, we use the fully nonlinear minimal flow unit simulations that resolve the minimum set of flow structures that sustain turbulence \citep{jimenez1991minimal}.  
First, we confirm if the trend obtained from the laminar flow-based model in figure \ref{fig:Amp_R1R2} is valid for a turbulent flow-based model. 
Figure \ref{fig:AmpTurb_R1R2} shows the predictions from the model \eqref{eqn:LNS_ViscStrat} that is now constructed using linearization around the turbulent mean profiles obtained from minimal flow unit simulations. 
Consistent with the laminar case, figure \ref{fig:AmpTurb_R1R2} shows that the model predicts streak enhancement at the more-viscous half of the Poiseuille flow; a trend that is not present for the Couette flow. 
At the low $Pr$ considered here, the asymmetry-induced modification of the classical lift-up mechanism drives this trend. 

\begin{figure}
\captionsetup[subfigure]{labelformat=empty,skip=-0pt}
\subfigure{
\centering
\centering
\includegraphics[width=0.9\textwidth]{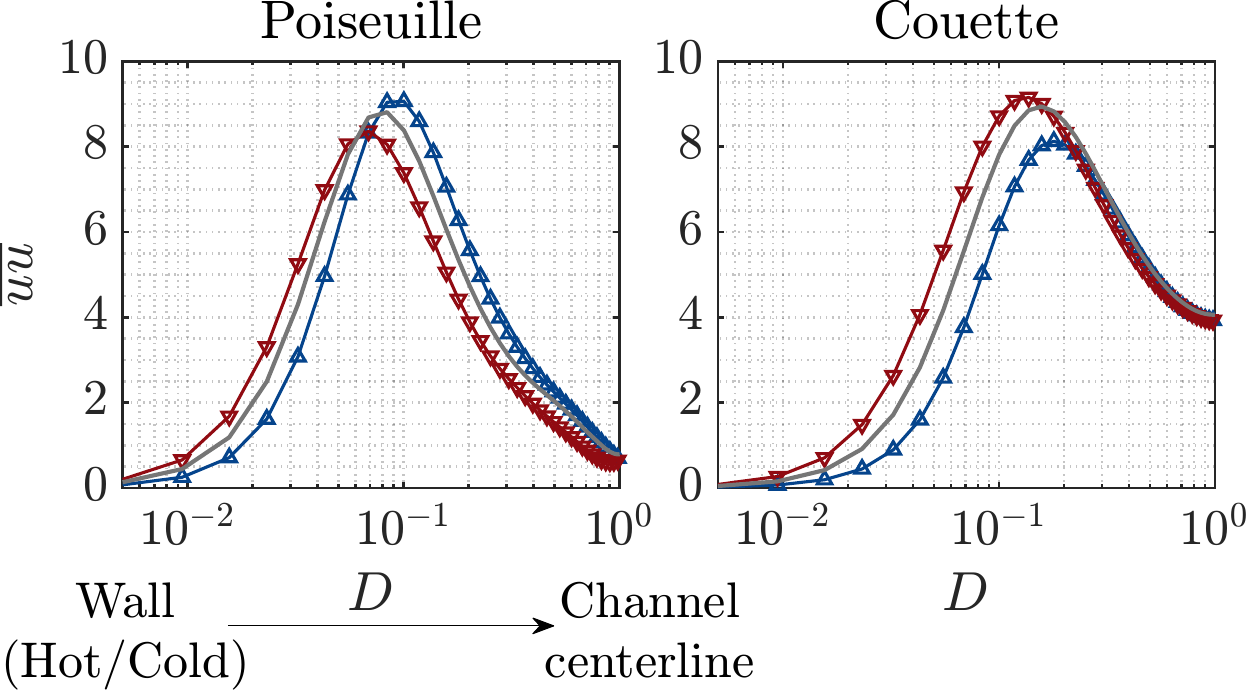}
\label{fig:MinChan_chan}
}%
\subfigure{
\centering
\label{fig:MinChan_couette}
}
\caption{
The variance of streamwise velocity $\overline{uu}$ obtained for $Pr=1$ and $\Delta\Theta^*=40$ minimal channel simulations. 
The more-viscous ($\triangle$ markers) and less-viscous ($\nabla$ markers) channel halves are compared against the unstratified channel half at the same Reynolds number (gray). 
Note, the profiles from the two channel halves are superimposed for ease of comparison, with $D$ denoting the distance from the walls (hot or cold). 
(\textit{a}) a Poiseuille flow at $Re_\tau=180$ (\textit{b}) a Couette flow at $Re_\tau=100$ are considered. 
}
\label{fig:MinChan}
\end{figure}

Now, to verify if this trend is present in the fully nonlinear flow, figure \ref{fig:MinChan}\subref{fig:MinChan_chan} shows the variance of streamwise velocity obtained from the simulation of a viscosity-stratified Poiseuille flow. 
In line with the theoretical observations here, and also consistent with the DNS of \citet{zonta2012modulation}, higher variance is obtained in the more-viscous half than in the less-viscous half. 
Turbulence is therefore enhanced in the more-viscous half. 
Now considering the Couette flow in figure \ref{fig:MinChan}\subref{fig:MinChan_couette}, contrary to the Poiseuille flow, higher variance is observed in the less-viscous half, i.e.\ turbulence is enhanced in the less-viscous half. 
The numerical experiments therefore support the theoretical findings presented here. 
It is also worth noting that, in figure \ref{fig:R1R2}, the model predicts a stratification-induced overall shift of $\widehat{u}$ towards the less-viscous region of the channel (i.e.\ shift towards top wall). 
In DNS, this is observed as a shift in the wall-normal location at which the variance peaks -- away from the wall in the more-viscous half and closer to the wall in the less-viscous half. 

The flow regimes considered so far span low $Pr$ where the modified lift-up is active, or moderate $Pr$ where the interaction of the classical and viscosity-stratified lift-up dictates the trends. 
The model predicts that, at higher values of $Pr$, the well-known classical lift-up mechanism will cease to impact the flow, and viscosity stratified lift-up will dominate.  
In these regimes, the length and time scales of the turbulent flow structures will therefore likely change, and this will impact practical applications like estimation and control.   
Additionally, viscosity stratification is one of several factors distinguishing streaks in compressible from incompressible flows. 
Even considering this single factor, the observations here suggest that the streaks in compressible flows are mechanistically distinct from their incompressible counterparts, therefore contributing to the ongoing debate on the equivalence of the streaks in these two flow regimes \citep[e.g.][]{smits1989comparison, pirozzoli2011turbulence, duan2011bdirect, williams2018experiments, bross2021large}. 
The observations here therefore highlight the importance of re-characterizing the physical mechanisms that generate the dominant features in flows where practically relevant factors like stratification are active. 

\mbox{}

\textit{Acknowledgments} -- AM and RG would like to thank the Isaac Newton Institute for Mathematical Sciences for support and hospitality during the program `Mathematical aspects of turbulence: where do we stand?' where part of the work on this paper was undertaken. 
This was partly supported by EPSRC grant number EP/R014604/1. 
AM gratefully acknowledges the financial support provided by the Department of Science and Technology (DST), Government of India, under the INSPIRE Faculty Fellowships, Grant number SP/DSTO-23-0192.  
RG thanks the Department of Atomic Energy, Government of India, for support under Project No. RTI4001.

\bibliography{ref}

\begin{thebibliography}{31}%
\makeatletter
\providecommand \@ifxundefined [1]{%
 \@ifx{#1\undefined}
}%
\providecommand \@ifnum [1]{%
 \ifnum #1\expandafter \@firstoftwo
 \else \expandafter \@secondoftwo
 \fi
}%
\providecommand \@ifx [1]{%
 \ifx #1\expandafter \@firstoftwo
 \else \expandafter \@secondoftwo
 \fi
}%
\providecommand \natexlab [1]{#1}%
\providecommand \enquote  [1]{``#1''}%
\providecommand \bibnamefont  [1]{#1}%
\providecommand \bibfnamefont [1]{#1}%
\providecommand \citenamefont [1]{#1}%
\providecommand \href@noop [0]{\@secondoftwo}%
\providecommand \href [0]{\begingroup \@sanitize@url \@href}%
\providecommand \@href[1]{\@@startlink{#1}\@@href}%
\providecommand \@@href[1]{\endgroup#1\@@endlink}%
\providecommand \@sanitize@url [0]{\catcode `\\12\catcode `\$12\catcode
  `\&12\catcode `\#12\catcode `\^12\catcode `\_12\catcode `\%12\relax}%
\providecommand \@@startlink[1]{}%
\providecommand \@@endlink[0]{}%
\providecommand \url  [0]{\begingroup\@sanitize@url \@url }%
\providecommand \@url [1]{\endgroup\@href {#1}{\urlprefix }}%
\providecommand \urlprefix  [0]{URL }%
\providecommand \Eprint [0]{\href }%
\providecommand \doibase [0]{https://doi.org/}%
\providecommand \selectlanguage [0]{\@gobble}%
\providecommand \bibinfo  [0]{\@secondoftwo}%
\providecommand \bibfield  [0]{\@secondoftwo}%
\providecommand \translation [1]{[#1]}%
\providecommand \BibitemOpen [0]{}%
\providecommand \bibitemStop [0]{}%
\providecommand \bibitemNoStop [0]{.\EOS\space}%
\providecommand \EOS [0]{\spacefactor3000\relax}%
\providecommand \BibitemShut  [1]{\csname bibitem#1\endcsname}%
\let\auto@bib@innerbib\@empty
\bibitem [{\citenamefont {Kline}\ \emph {et~al.}(1967)\citenamefont {Kline},
  \citenamefont {Reynolds}, \citenamefont {Schraub},\ and\ \citenamefont
  {Runstadler}}]{kline1967structure}%
  \BibitemOpen
  \bibfield  {author} {\bibinfo {author} {\bibfnamefont {S.~J.}\ \bibnamefont
  {Kline}}, \bibinfo {author} {\bibfnamefont {W.~C.}\ \bibnamefont {Reynolds}},
  \bibinfo {author} {\bibfnamefont {F.~A.}\ \bibnamefont {Schraub}},\ and\
  \bibinfo {author} {\bibfnamefont {P.~W.}\ \bibnamefont {Runstadler}},\
  }\bibfield  {title} {\bibinfo {title} {The structure of turbulent boundary
  layers},\ }\href@noop {} {\bibfield  {journal} {\bibinfo  {journal} {J. Fluid
  Mech.}\ }\textbf {\bibinfo {volume} {30}},\ \bibinfo {pages} {741} (\bibinfo
  {year} {1967})}\BibitemShut {NoStop}%
\bibitem [{\citenamefont {Smits}\ \emph {et~al.}(2011)\citenamefont {Smits},
  \citenamefont {McKeon},\ and\ \citenamefont {Marusic}}]{smits2011high}%
  \BibitemOpen
  \bibfield  {author} {\bibinfo {author} {\bibfnamefont {A.~J.}\ \bibnamefont
  {Smits}}, \bibinfo {author} {\bibfnamefont {B.~J.}\ \bibnamefont {McKeon}},\
  and\ \bibinfo {author} {\bibfnamefont {I.}~\bibnamefont {Marusic}},\
  }\bibfield  {title} {\bibinfo {title} {High-{R}eynolds number wall
  turbulence},\ }\href@noop {} {\bibfield  {journal} {\bibinfo  {journal}
  {Annu. Rev. Fluid Mech.}\ }\textbf {\bibinfo {volume} {43}},\ \bibinfo
  {pages} {353} (\bibinfo {year} {2011})}\BibitemShut {NoStop}%
\bibitem [{\citenamefont {K{\"u}hnen}\ \emph {et~al.}(2018)\citenamefont
  {K{\"u}hnen}, \citenamefont {Song}, \citenamefont {Scarselli}, \citenamefont
  {Budanur}, \citenamefont {Riedl}, \citenamefont {Willis}, \citenamefont
  {Avila},\ and\ \citenamefont {Hof}}]{kuhnen2018destabilizing}%
  \BibitemOpen
  \bibfield  {author} {\bibinfo {author} {\bibfnamefont {J.}~\bibnamefont
  {K{\"u}hnen}}, \bibinfo {author} {\bibfnamefont {B.}~\bibnamefont {Song}},
  \bibinfo {author} {\bibfnamefont {D.}~\bibnamefont {Scarselli}}, \bibinfo
  {author} {\bibfnamefont {N.~B.}\ \bibnamefont {Budanur}}, \bibinfo {author}
  {\bibfnamefont {M.}~\bibnamefont {Riedl}}, \bibinfo {author} {\bibfnamefont
  {A.~P.}\ \bibnamefont {Willis}}, \bibinfo {author} {\bibfnamefont
  {M.}~\bibnamefont {Avila}},\ and\ \bibinfo {author} {\bibfnamefont
  {B.}~\bibnamefont {Hof}},\ }\bibfield  {title} {\bibinfo {title}
  {Destabilizing turbulence in pipe flow},\ }\href@noop {} {\bibfield
  {journal} {\bibinfo  {journal} {Nat. Phys.}\ }\textbf {\bibinfo {volume}
  {14}},\ \bibinfo {pages} {386} (\bibinfo {year} {2018})}\BibitemShut
  {NoStop}%
\bibitem [{\citenamefont {Marusic}\ \emph {et~al.}(2021)\citenamefont
  {Marusic}, \citenamefont {Chandran}, \citenamefont {Rouhi}, \citenamefont
  {Fu}, \citenamefont {Wine}, \citenamefont {Holloway}, \citenamefont {Chung},\
  and\ \citenamefont {Smits}}]{marusic2021energy}%
  \BibitemOpen
  \bibfield  {author} {\bibinfo {author} {\bibfnamefont {I.}~\bibnamefont
  {Marusic}}, \bibinfo {author} {\bibfnamefont {D.}~\bibnamefont {Chandran}},
  \bibinfo {author} {\bibfnamefont {A.}~\bibnamefont {Rouhi}}, \bibinfo
  {author} {\bibfnamefont {M.~K.}\ \bibnamefont {Fu}}, \bibinfo {author}
  {\bibfnamefont {D.}~\bibnamefont {Wine}}, \bibinfo {author} {\bibfnamefont
  {B.}~\bibnamefont {Holloway}}, \bibinfo {author} {\bibfnamefont
  {D.}~\bibnamefont {Chung}},\ and\ \bibinfo {author} {\bibfnamefont {A.~J.}\
  \bibnamefont {Smits}},\ }\bibfield  {title} {\bibinfo {title} {An
  energy-efficient pathway to turbulent drag reduction},\ }\href@noop {}
  {\bibfield  {journal} {\bibinfo  {journal} {Nat. Commun.}\ }\textbf {\bibinfo
  {volume} {12}},\ \bibinfo {pages} {5805} (\bibinfo {year}
  {2021})}\BibitemShut {NoStop}%
\bibitem [{\citenamefont {Hutchins}\ \emph {et~al.}(2012)\citenamefont
  {Hutchins}, \citenamefont {Chauhan}, \citenamefont {Marusic}, \citenamefont
  {Monty},\ and\ \citenamefont {Klewicki}}]{hutchins2012towards}%
  \BibitemOpen
  \bibfield  {author} {\bibinfo {author} {\bibfnamefont {N.}~\bibnamefont
  {Hutchins}}, \bibinfo {author} {\bibfnamefont {K.}~\bibnamefont {Chauhan}},
  \bibinfo {author} {\bibfnamefont {I.}~\bibnamefont {Marusic}}, \bibinfo
  {author} {\bibfnamefont {J.}~\bibnamefont {Monty}},\ and\ \bibinfo {author}
  {\bibfnamefont {J.}~\bibnamefont {Klewicki}},\ }\bibfield  {title} {\bibinfo
  {title} {Towards reconciling the large-scale structure of turbulent boundary
  layers in the atmosphere and laboratory},\ }\href@noop {} {\bibfield
  {journal} {\bibinfo  {journal} {Bound.-Layer Meteorol.}\ }\textbf {\bibinfo
  {volume} {145}},\ \bibinfo {pages} {273} (\bibinfo {year}
  {2012})}\BibitemShut {NoStop}%
\bibitem [{\citenamefont {Momen}\ \emph {et~al.}(2021)\citenamefont {Momen},
  \citenamefont {Parlange},\ and\ \citenamefont
  {Giometto}}]{momen2021scrambling}%
  \BibitemOpen
  \bibfield  {author} {\bibinfo {author} {\bibfnamefont {M.}~\bibnamefont
  {Momen}}, \bibinfo {author} {\bibfnamefont {M.~B.}\ \bibnamefont
  {Parlange}},\ and\ \bibinfo {author} {\bibfnamefont {M.~G.}\ \bibnamefont
  {Giometto}},\ }\bibfield  {title} {\bibinfo {title} {Scrambling and
  reorientation of classical atmospheric boundary layer turbulence in hurricane
  winds},\ }\href@noop {} {\bibfield  {journal} {\bibinfo  {journal} {Geophys.
  Res. Lett.}\ }\textbf {\bibinfo {volume} {48}},\ \bibinfo {pages}
  {e2020GL091695} (\bibinfo {year} {2021})}\BibitemShut {NoStop}%
\bibitem [{\citenamefont {Landahl}(1980)}]{landahl1980note}%
  \BibitemOpen
  \bibfield  {author} {\bibinfo {author} {\bibfnamefont {M.~T.}\ \bibnamefont
  {Landahl}},\ }\bibfield  {title} {\bibinfo {title} {A note on an algebraic
  instability of inviscid parallel shear flows},\ }\href@noop {} {\bibfield
  {journal} {\bibinfo  {journal} {J. Fluid Mech.}\ }\textbf {\bibinfo {volume}
  {98}},\ \bibinfo {pages} {243} (\bibinfo {year} {1980})}\BibitemShut
  {NoStop}%
\bibitem [{\citenamefont {Ellingsen}\ and\ \citenamefont
  {Palm}(1975)}]{ellingsen1975stability}%
  \BibitemOpen
  \bibfield  {author} {\bibinfo {author} {\bibfnamefont {T.}~\bibnamefont
  {Ellingsen}}\ and\ \bibinfo {author} {\bibfnamefont {E.}~\bibnamefont
  {Palm}},\ }\bibfield  {title} {\bibinfo {title} {Stability of linear flow},\
  }\href@noop {} {\bibfield  {journal} {\bibinfo  {journal} {Phys. Fluids}\
  }\textbf {\bibinfo {volume} {18}},\ \bibinfo {pages} {487} (\bibinfo {year}
  {1975})}\BibitemShut {NoStop}%
\bibitem [{\citenamefont {Steffen}\ and\ \citenamefont
  {Kaufmann}(2005)}]{steffen2005glacial}%
  \BibitemOpen
  \bibfield  {author} {\bibinfo {author} {\bibfnamefont {H.}~\bibnamefont
  {Steffen}}\ and\ \bibinfo {author} {\bibfnamefont {G.}~\bibnamefont
  {Kaufmann}},\ }\bibfield  {title} {\bibinfo {title} {Glacial isostatic
  adjustment of {S}candinavia and northwestern {E}urope and the radial
  viscosity structure of the earth's mantle},\ }\href@noop {} {\bibfield
  {journal} {\bibinfo  {journal} {Geophys. J. Int.}\ }\textbf {\bibinfo
  {volume} {163}},\ \bibinfo {pages} {801} (\bibinfo {year}
  {2005})}\BibitemShut {NoStop}%
\bibitem [{\citenamefont {Beal}(1946)}]{beal1946viscosity}%
  \BibitemOpen
  \bibfield  {author} {\bibinfo {author} {\bibfnamefont {C.}~\bibnamefont
  {Beal}},\ }\bibfield  {title} {\bibinfo {title} {The viscosity of air, water,
  natural gas, crude oil and its associated gases at oil field temperatures and
  pressures},\ }\href@noop {} {\bibfield  {journal} {\bibinfo  {journal}
  {Trans. AIME}\ }\textbf {\bibinfo {volume} {165}},\ \bibinfo {pages} {94}
  (\bibinfo {year} {1946})}\BibitemShut {NoStop}%
\bibitem [{\citenamefont {Kalashnikov}(1994)}]{kalashnikov1994shear}%
  \BibitemOpen
  \bibfield  {author} {\bibinfo {author} {\bibfnamefont {V.~N.}\ \bibnamefont
  {Kalashnikov}},\ }\bibfield  {title} {\bibinfo {title} {Shear-rate dependent
  viscosity of dilute polymer solutions},\ }\href@noop {} {\bibfield  {journal}
  {\bibinfo  {journal} {J. Rheol.}\ }\textbf {\bibinfo {volume} {38}},\
  \bibinfo {pages} {1385} (\bibinfo {year} {1994})}\BibitemShut {NoStop}%
\bibitem [{\citenamefont {Wilke}(1950)}]{wilke1950viscosity}%
  \BibitemOpen
  \bibfield  {author} {\bibinfo {author} {\bibfnamefont {C.~R.}\ \bibnamefont
  {Wilke}},\ }\bibfield  {title} {\bibinfo {title} {A viscosity equation for
  gas mixtures},\ }\href@noop {} {\bibfield  {journal} {\bibinfo  {journal} {J.
  Chem. Phys.}\ }\textbf {\bibinfo {volume} {18}},\ \bibinfo {pages} {517}
  (\bibinfo {year} {1950})}\BibitemShut {NoStop}%
\bibitem [{\citenamefont {Govindarajan}\ \emph {et~al.}(2001)\citenamefont
  {Govindarajan}, \citenamefont {L'vov},\ and\ \citenamefont
  {Procaccia}}]{govindarajan2001retardation}%
  \BibitemOpen
  \bibfield  {author} {\bibinfo {author} {\bibfnamefont {R.}~\bibnamefont
  {Govindarajan}}, \bibinfo {author} {\bibfnamefont {V.~S.}\ \bibnamefont
  {L'vov}},\ and\ \bibinfo {author} {\bibfnamefont {I.}~\bibnamefont
  {Procaccia}},\ }\bibfield  {title} {\bibinfo {title} {Retardation of the
  onset of turbulence by minor viscosity contrasts},\ }\href@noop {} {\bibfield
   {journal} {\bibinfo  {journal} {Phys. Rev. Lett.}\ }\textbf {\bibinfo
  {volume} {87}},\ \bibinfo {pages} {174501} (\bibinfo {year}
  {2001})}\BibitemShut {NoStop}%
\bibitem [{\citenamefont {Chikkadi}\ \emph {et~al.}(2005)\citenamefont
  {Chikkadi}, \citenamefont {Sameen},\ and\ \citenamefont
  {Govindarajan}}]{chikkadi2005preventing}%
  \BibitemOpen
  \bibfield  {author} {\bibinfo {author} {\bibfnamefont {V.}~\bibnamefont
  {Chikkadi}}, \bibinfo {author} {\bibfnamefont {A.}~\bibnamefont {Sameen}},\
  and\ \bibinfo {author} {\bibfnamefont {R.}~\bibnamefont {Govindarajan}},\
  }\bibfield  {title} {\bibinfo {title} {Preventing transition to turbulence: A
  viscosity stratification does not always help},\ }\href@noop {} {\bibfield
  {journal} {\bibinfo  {journal} {Phys. Rev. Lett.}\ }\textbf {\bibinfo
  {volume} {95}},\ \bibinfo {pages} {264504} (\bibinfo {year}
  {2005})}\BibitemShut {NoStop}%
\bibitem [{\citenamefont {Govindarajan}\ and\ \citenamefont
  {Sahu}(2014)}]{govindarajan2014instabilities}%
  \BibitemOpen
  \bibfield  {author} {\bibinfo {author} {\bibfnamefont {R.}~\bibnamefont
  {Govindarajan}}\ and\ \bibinfo {author} {\bibfnamefont {K.~C.}\ \bibnamefont
  {Sahu}},\ }\bibfield  {title} {\bibinfo {title} {Instabilities in
  viscosity-stratified flow},\ }\href@noop {} {\bibfield  {journal} {\bibinfo
  {journal} {Annu. Rev. Fluid Mech.}\ }\textbf {\bibinfo {volume} {46}},\
  \bibinfo {pages} {331} (\bibinfo {year} {2014})}\BibitemShut {NoStop}%
\bibitem [{\citenamefont {Thakur}\ \emph {et~al.}(2021)\citenamefont {Thakur},
  \citenamefont {Sharma},\ and\ \citenamefont
  {Govindarajan}}]{thakur2021early}%
  \BibitemOpen
  \bibfield  {author} {\bibinfo {author} {\bibfnamefont {R.}~\bibnamefont
  {Thakur}}, \bibinfo {author} {\bibfnamefont {A.}~\bibnamefont {Sharma}},\
  and\ \bibinfo {author} {\bibfnamefont {R.}~\bibnamefont {Govindarajan}},\
  }\bibfield  {title} {\bibinfo {title} {Early evolution of optimal
  perturbations in a viscosity-stratified channel},\ }\href@noop {} {\bibfield
  {journal} {\bibinfo  {journal} {J. Fluid Mech.}\ }\textbf {\bibinfo {volume}
  {914}},\ \bibinfo {pages} {A10} (\bibinfo {year} {2021})}\BibitemShut
  {NoStop}%
\bibitem [{\citenamefont {Rinaldi}\ \emph {et~al.}(2018)\citenamefont
  {Rinaldi}, \citenamefont {Schlatter},\ and\ \citenamefont
  {Bagheri}}]{rinaldi2018edge}%
  \BibitemOpen
  \bibfield  {author} {\bibinfo {author} {\bibfnamefont {E.}~\bibnamefont
  {Rinaldi}}, \bibinfo {author} {\bibfnamefont {P.}~\bibnamefont {Schlatter}},\
  and\ \bibinfo {author} {\bibfnamefont {S.}~\bibnamefont {Bagheri}},\
  }\bibfield  {title} {\bibinfo {title} {Edge state modulation by mean
  viscosity gradients},\ }\href@noop {} {\bibfield  {journal} {\bibinfo
  {journal} {J. Fluid Mech.}\ }\textbf {\bibinfo {volume} {838}},\ \bibinfo
  {pages} {379} (\bibinfo {year} {2018})}\BibitemShut {NoStop}%
\bibitem [{\citenamefont {Lee}\ \emph {et~al.}(2013)\citenamefont {Lee},
  \citenamefont {Jung}, \citenamefont {Sung},\ and\ \citenamefont
  {Zaki}}]{lee2013effect}%
  \BibitemOpen
  \bibfield  {author} {\bibinfo {author} {\bibfnamefont {J.}~\bibnamefont
  {Lee}}, \bibinfo {author} {\bibfnamefont {S.~Y.}\ \bibnamefont {Jung}},
  \bibinfo {author} {\bibfnamefont {H.~J.}\ \bibnamefont {Sung}},\ and\
  \bibinfo {author} {\bibfnamefont {T.~A.}\ \bibnamefont {Zaki}},\ }\bibfield
  {title} {\bibinfo {title} {Effect of wall heating on turbulent boundary
  layers with temperature-dependent viscosity},\ }\href@noop {} {\bibfield
  {journal} {\bibinfo  {journal} {J. Fluid Mech.}\ }\textbf {\bibinfo {volume}
  {726}},\ \bibinfo {pages} {196} (\bibinfo {year} {2013})}\BibitemShut
  {NoStop}%
\bibitem [{\citenamefont {Zonta}\ \emph {et~al.}(2012)\citenamefont {Zonta},
  \citenamefont {Marchioli},\ and\ \citenamefont
  {Soldati}}]{zonta2012modulation}%
  \BibitemOpen
  \bibfield  {author} {\bibinfo {author} {\bibfnamefont {F.}~\bibnamefont
  {Zonta}}, \bibinfo {author} {\bibfnamefont {C.}~\bibnamefont {Marchioli}},\
  and\ \bibinfo {author} {\bibfnamefont {A.}~\bibnamefont {Soldati}},\
  }\bibfield  {title} {\bibinfo {title} {Modulation of turbulence in forced
  convection by temperature-dependent viscosity},\ }\href@noop {} {\bibfield
  {journal} {\bibinfo  {journal} {J. Fluid Mech.}\ }\textbf {\bibinfo {volume}
  {697}},\ \bibinfo {pages} {150} (\bibinfo {year} {2012})}\BibitemShut
  {NoStop}%
\bibitem [{\citenamefont {Jovanovi{\'c}}\ and\ \citenamefont
  {Bamieh}(2005)}]{jovanovic2005componentwise}%
  \BibitemOpen
  \bibfield  {author} {\bibinfo {author} {\bibfnamefont {M.~R.}\ \bibnamefont
  {Jovanovi{\'c}}}\ and\ \bibinfo {author} {\bibfnamefont {B.}~\bibnamefont
  {Bamieh}},\ }\bibfield  {title} {\bibinfo {title} {Componentwise energy
  amplification in channel flows},\ }\href@noop {} {\bibfield  {journal}
  {\bibinfo  {journal} {J. Fluid Mech.}\ }\textbf {\bibinfo {volume} {534}},\
  \bibinfo {pages} {145} (\bibinfo {year} {2005})}\BibitemShut {NoStop}%
\bibitem [{\citenamefont {Illingworth}\ \emph {et~al.}(2018)\citenamefont
  {Illingworth}, \citenamefont {Monty},\ and\ \citenamefont
  {Marusic}}]{illingworth2018estimating}%
  \BibitemOpen
  \bibfield  {author} {\bibinfo {author} {\bibfnamefont {S.~J.}\ \bibnamefont
  {Illingworth}}, \bibinfo {author} {\bibfnamefont {J.~P.}\ \bibnamefont
  {Monty}},\ and\ \bibinfo {author} {\bibfnamefont {I.}~\bibnamefont
  {Marusic}},\ }\bibfield  {title} {\bibinfo {title} {Estimating large-scale
  structures in wall turbulence using linear models},\ }\href@noop {}
  {\bibfield  {journal} {\bibinfo  {journal} {J. Fluid Mech.}\ }\textbf
  {\bibinfo {volume} {842}},\ \bibinfo {pages} {146} (\bibinfo {year}
  {2018})}\BibitemShut {NoStop}%
\bibitem [{\citenamefont {Illingworth}(2020)}]{illingworth2020streamwise}%
  \BibitemOpen
  \bibfield  {author} {\bibinfo {author} {\bibfnamefont {S.~J.}\ \bibnamefont
  {Illingworth}},\ }\bibfield  {title} {\bibinfo {title} {Streamwise-constant
  large-scale structures in {C}ouette and {P}oiseuille flows},\ }\href@noop {}
  {\bibfield  {journal} {\bibinfo  {journal} {J. Fluid Mech.}\ }\textbf
  {\bibinfo {volume} {889}},\ \bibinfo {pages} {A13} (\bibinfo {year}
  {2020})}\BibitemShut {NoStop}%
\bibitem [{\citenamefont {Jim{\'e}nez}\ and\ \citenamefont
  {Moin}(1991)}]{jimenez1991minimal}%
  \BibitemOpen
  \bibfield  {author} {\bibinfo {author} {\bibfnamefont {J.}~\bibnamefont
  {Jim{\'e}nez}}\ and\ \bibinfo {author} {\bibfnamefont {P.}~\bibnamefont
  {Moin}},\ }\bibfield  {title} {\bibinfo {title} {The minimal flow unit in
  near-wall turbulence},\ }\href@noop {} {\bibfield  {journal} {\bibinfo
  {journal} {J. Fluid Mech.}\ }\textbf {\bibinfo {volume} {225}},\ \bibinfo
  {pages} {213} (\bibinfo {year} {1991})}\BibitemShut {NoStop}%
\bibitem [{\citenamefont {Smits}\ \emph {et~al.}(1989)\citenamefont {Smits},
  \citenamefont {Spina}, \citenamefont {Alving}, \citenamefont {Smith},
  \citenamefont {Fernando},\ and\ \citenamefont
  {Donovan}}]{smits1989comparison}%
  \BibitemOpen
  \bibfield  {author} {\bibinfo {author} {\bibfnamefont {A.~J.}\ \bibnamefont
  {Smits}}, \bibinfo {author} {\bibfnamefont {E.~F.}\ \bibnamefont {Spina}},
  \bibinfo {author} {\bibfnamefont {A.~E.}\ \bibnamefont {Alving}}, \bibinfo
  {author} {\bibfnamefont {R.~W.}\ \bibnamefont {Smith}}, \bibinfo {author}
  {\bibfnamefont {E.~M.}\ \bibnamefont {Fernando}},\ and\ \bibinfo {author}
  {\bibfnamefont {J.~F.}\ \bibnamefont {Donovan}},\ }\bibfield  {title}
  {\bibinfo {title} {A comparison of the turbulence structure of subsonic and
  supersonic boundary layers},\ }\href@noop {} {\bibfield  {journal} {\bibinfo
  {journal} {Phys. Fluids A}\ }\textbf {\bibinfo {volume} {1}},\ \bibinfo
  {pages} {1865} (\bibinfo {year} {1989})}\BibitemShut {NoStop}%
\bibitem [{\citenamefont {Pirozzoli}\ and\ \citenamefont
  {Bernardini}(2011)}]{pirozzoli2011turbulence}%
  \BibitemOpen
  \bibfield  {author} {\bibinfo {author} {\bibfnamefont {S.}~\bibnamefont
  {Pirozzoli}}\ and\ \bibinfo {author} {\bibfnamefont {M.}~\bibnamefont
  {Bernardini}},\ }\bibfield  {title} {\bibinfo {title} {Turbulence in
  supersonic boundary layers at moderate {R}eynolds number},\ }\href@noop {}
  {\bibfield  {journal} {\bibinfo  {journal} {J. Fluid Mech.}\ }\textbf
  {\bibinfo {volume} {688}},\ \bibinfo {pages} {120} (\bibinfo {year}
  {2011})}\BibitemShut {NoStop}%
\bibitem [{\citenamefont {Duan}\ and\ \citenamefont
  {Martin}(2011)}]{duan2011bdirect}%
  \BibitemOpen
  \bibfield  {author} {\bibinfo {author} {\bibfnamefont {L.}~\bibnamefont
  {Duan}}\ and\ \bibinfo {author} {\bibfnamefont {M.~P.}\ \bibnamefont
  {Martin}},\ }\bibfield  {title} {\bibinfo {title} {Direct numerical
  simulation of hypersonic turbulent boundary layers. {P}art 4. {E}ffect of
  high enthalpy},\ }\href@noop {} {\bibfield  {journal} {\bibinfo  {journal}
  {J. Fluid Mech.}\ }\textbf {\bibinfo {volume} {684}},\ \bibinfo {pages} {25}
  (\bibinfo {year} {2011})}\BibitemShut {NoStop}%
\bibitem [{\citenamefont {Williams}\ \emph {et~al.}(2018)\citenamefont
  {Williams}, \citenamefont {Sahoo}, \citenamefont {Baumgartner},\ and\
  \citenamefont {Smits}}]{williams2018experiments}%
  \BibitemOpen
  \bibfield  {author} {\bibinfo {author} {\bibfnamefont {O.~J.~H.}\
  \bibnamefont {Williams}}, \bibinfo {author} {\bibfnamefont {D.}~\bibnamefont
  {Sahoo}}, \bibinfo {author} {\bibfnamefont {M.~L.}\ \bibnamefont
  {Baumgartner}},\ and\ \bibinfo {author} {\bibfnamefont {A.~J.}\ \bibnamefont
  {Smits}},\ }\bibfield  {title} {\bibinfo {title} {Experiments on the
  structure and scaling of hypersonic turbulent boundary layers},\ }\href@noop
  {} {\bibfield  {journal} {\bibinfo  {journal} {J. Fluid Mech.}\ }\textbf
  {\bibinfo {volume} {834}},\ \bibinfo {pages} {237} (\bibinfo {year}
  {2018})}\BibitemShut {NoStop}%
\bibitem [{\citenamefont {Bross}\ \emph {et~al.}(2021)\citenamefont {Bross},
  \citenamefont {Scharnowski},\ and\ \citenamefont
  {K{\"a}hler}}]{bross2021large}%
  \BibitemOpen
  \bibfield  {author} {\bibinfo {author} {\bibfnamefont {M.}~\bibnamefont
  {Bross}}, \bibinfo {author} {\bibfnamefont {S.}~\bibnamefont {Scharnowski}},\
  and\ \bibinfo {author} {\bibfnamefont {C.~J.}\ \bibnamefont {K{\"a}hler}},\
  }\bibfield  {title} {\bibinfo {title} {Large-scale coherent structures in
  compressible turbulent boundary layers},\ }\href@noop {} {\bibfield
  {journal} {\bibinfo  {journal} {J. Fluid Mech.}\ }\textbf {\bibinfo {volume}
  {911}} (\bibinfo {year} {2021})}\BibitemShut {NoStop}%
\bibitem [{\citenamefont {Burns}\ \emph {et~al.}(2020)\citenamefont {Burns},
  \citenamefont {Vasil}, \citenamefont {Oishi}, \citenamefont {Lecoanet},\ and\
  \citenamefont {Brown}}]{burns2020dedalus}%
  \BibitemOpen
  \bibfield  {author} {\bibinfo {author} {\bibfnamefont {K.~J.}\ \bibnamefont
  {Burns}}, \bibinfo {author} {\bibfnamefont {G.~M.}\ \bibnamefont {Vasil}},
  \bibinfo {author} {\bibfnamefont {J.~S.}\ \bibnamefont {Oishi}}, \bibinfo
  {author} {\bibfnamefont {D.}~\bibnamefont {Lecoanet}},\ and\ \bibinfo
  {author} {\bibfnamefont {B.~P.}\ \bibnamefont {Brown}},\ }\bibfield  {title}
  {\bibinfo {title} {Dedalus: {A} flexible framework for numerical simulations
  with spectral methods},\ }\href@noop {} {\bibfield  {journal} {\bibinfo
  {journal} {Phys. Rev. Res.}\ }\textbf {\bibinfo {volume} {2}},\ \bibinfo
  {pages} {023068} (\bibinfo {year} {2020})}\BibitemShut {NoStop}%
\bibitem [{\citenamefont {Bae}\ \emph {et~al.}(2021)\citenamefont {Bae},
  \citenamefont {Lozano-Duran},\ and\ \citenamefont
  {McKeon}}]{bae2021nonlinear}%
  \BibitemOpen
  \bibfield  {author} {\bibinfo {author} {\bibfnamefont {H.~J.}\ \bibnamefont
  {Bae}}, \bibinfo {author} {\bibfnamefont {A.}~\bibnamefont {Lozano-Duran}},\
  and\ \bibinfo {author} {\bibfnamefont {B.~J.}\ \bibnamefont {McKeon}},\
  }\bibfield  {title} {\bibinfo {title} {Nonlinear mechanism of the
  self-sustaining process in the buffer and logarithmic layer of wall-bounded
  flows},\ }\href@noop {} {\bibfield  {journal} {\bibinfo  {journal} {J. Fluid
  Mech.}\ }\textbf {\bibinfo {volume} {914}},\ \bibinfo {pages} {A3} (\bibinfo
  {year} {2021})}\BibitemShut {NoStop}%
\bibitem [{\citenamefont {Hamilton}\ \emph {et~al.}(1995)\citenamefont
  {Hamilton}, \citenamefont {Kim},\ and\ \citenamefont
  {Waleffe}}]{hamilton1995regeneration}%
  \BibitemOpen
  \bibfield  {author} {\bibinfo {author} {\bibfnamefont {J.~M.}\ \bibnamefont
  {Hamilton}}, \bibinfo {author} {\bibfnamefont {J.}~\bibnamefont {Kim}},\ and\
  \bibinfo {author} {\bibfnamefont {F.}~\bibnamefont {Waleffe}},\ }\bibfield
  {title} {\bibinfo {title} {Regeneration mechanisms of near-wall turbulence
  structures},\ }\href@noop {} {\bibfield  {journal} {\bibinfo  {journal} {J.
  Fluid Mech.}\ }\textbf {\bibinfo {volume} {287}},\ \bibinfo {pages} {317}
  (\bibinfo {year} {1995})}\BibitemShut {NoStop}%
\end{thebibliography}%

\appendix

\begin{onecolumngrid}
\vspace{0.5cm}  
\section*{END MATTER}
\end{onecolumngrid}

\begin{twocolumngrid}

\textbf{\textit{Governing equations}}: 
For the linear model in \eqref{eqn:OrrSommerfeldSquire}, velocities are  non-dimensionalized by the bulk velocity $U_b^*=0.5\int_{-1}^{1}U^*(z) dz$, temperature by $\Theta_{ref}^*=(\Theta_H^*+\Theta_C^*)/2$, the spatial variables by the channel half-height $h^*$, and the viscosity by the viscosity at the reference temperature $\mu_{ref}^*$.  
The non-dimensional Navier--Stokes equations linearized around the mean flow are:
\begin{equation}
\begin{split}
\frac{\partial u_i }{\partial t} &= 
- U_j \frac{\partial u_i}{\partial x_j} - u_j \frac{\partial U_i}{\partial x_j}
- \frac{\partial p}{\partial x_i} + \\
&\frac{1}{Re} \frac{\partial }{\partial x_j} \left( 
\mu(\Theta) \left( \frac{\partial u_i}{\partial x_j} + \frac{\partial u_j}{\partial x_i} \right) 
\right. \\ &\left.
+ \frac{\partial \mu(\Theta)}{\partial \Theta} \left( \frac{\partial U_i}{\partial x_j} + \frac{\partial U_j}{\partial x_i} \right) \theta  \right) + f_i, \qquad 
\frac{\partial u_i }{\partial x_i} = 0, \\
\frac{\partial \theta}{\partial t} &= 
- U_j \frac{\partial \theta}{\partial x_j} 
- u_j \frac{\partial \Theta}{\partial x_j} 
+ \frac{1}{PrRe} \frac{\partial^2 \theta}{\partial x_j\partial x_j} + f_\theta. \\
\end{split}
\label{eqn:LNS_ViscStrat}
\end{equation}
Here $(x_1,x_2,x_3)=(x,y,z)$ and $(u_1,u_2,u_3)=(u,v,w)$ and $p$ is pressure. 
Viscosity is linearized as $\mu(\Theta+\theta) = \mu(\Theta) + (d\mu(\Theta)/d\Theta)\theta$, where $\mu(\Theta)$ is the mean viscosity and $(d\mu(\Theta)/d\Theta)\theta$ are the fluctuations of viscosity.  

\begin{figure}
\subfigure{
\centering
\includegraphics[width=\textwidth]{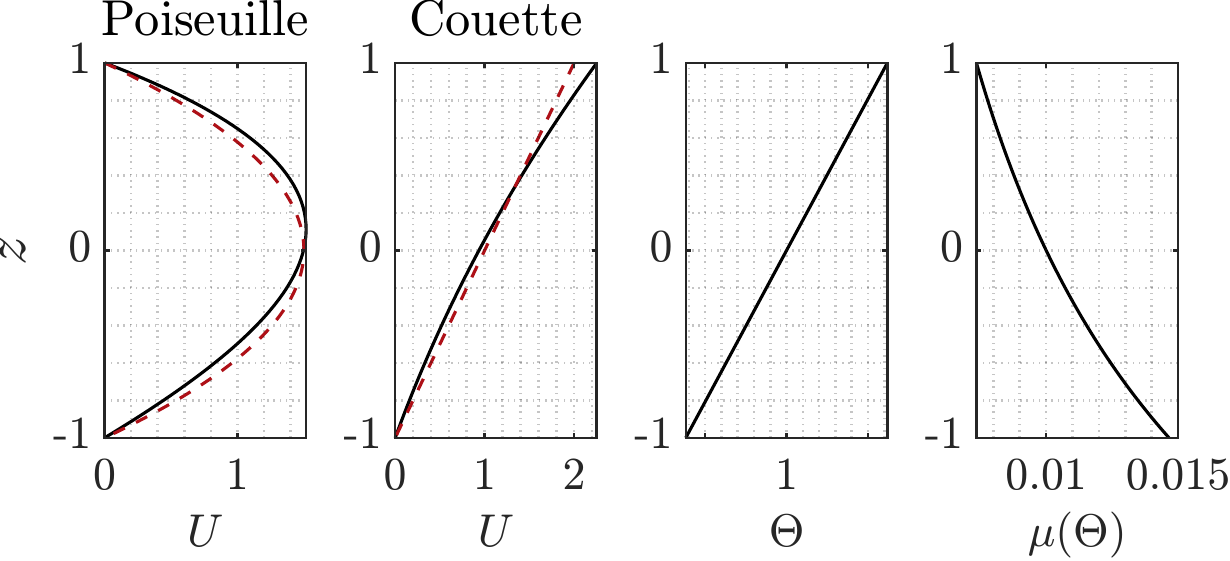}
\label{fig:laminar_MeanProfiles}
}
\caption{
Mean velocity profiles for laminar (\textit{a}) Poiseuille and (\textit{b}) Couette flows with $\Delta\Theta^* = 40K$ (black, solid) and $\Delta\Theta^* = 0K$, i.e.\ unstratified (red, dashed). 
The mean (\textit{c}) temperature and (\textit{d}) viscosity profiles are also shown.}
\label{fig:MeanProfiles}
\end{figure}
The laminar mean profiles that are provided as input to \eqref{eqn:LNS_ViscStrat} are obtained by first assuming a mean temperature that varies linearly with $z$. 
To obtain the mean velocities for the Poiseuille flow, the mean momentum equations are solved assuming the mean imposed pressure gradient $d\overline{P}/dx$ to be constant for the unstratified and stratified flows. 
Similarly, the mean velocity profiles for the Couette flow are obtained by assuming a constant velocity for the top wall. 
The laminar mean profiles do not change with $Re$ or $Pr$ and are shown in figure \ref{fig:MeanProfiles}. 
Fourier transforming in $x$, $y$ and time and considering the $k_x=0$ structures alone in \eqref{eqn:LNS_ViscStrat} gives \eqref{eqn:OrrSommerfeldSquire}. 
The wavenumbers ($k_x,k_y$) are non-dimesionalized by $1/h^*$. 

\textbf{\textit{Minimal channel simulations}}: The minimal channel simulations reported in this study are carried out using the open source partial differential equation solver Dedalus \citep{burns2020dedalus}. 
Fourier decompositions are used in the $x$ and $y$ directions and a Chebyshev decomposition is employed in the $z$ direction. 
Viscosity varies with temperature, as in \eqref{eqn:visc}. 
The simulations are performed at a constant friction Reynolds number $Re_\tau=\rho_{ref}^* u_\tau^* h^*/\mu_{ref}^*$, where $u_\tau^* = (\tau_w^*/\rho_{ref}^*)$ is the friction velocity defined using the wall-shear stress $\tau_w^*$. 
In viscosity-stratified channel flows, the wall-shear stress at the hot and the cold walls are different. 
We therefore use $\tau_w^* = (|\tau_{w,C}^*| + |\tau_{w,H}^*|)/2$ \citep{zonta2012modulation}. 
The spatial variables are normalized by $h^*$ and the velocities by $u_\tau^*$. 
The Prandtl number is kept at $Pr=1$ for low computational cost. 
We consider $\Delta\Theta^* = 40K$. 

The full Navier-Stokes equations are solved \citep{zonta2012modulation}. 
For the Poiseuille flow, we impose a pressure gradient of $-1$ and use Dirichlet velocity boundary conditions at both the walls, and for the Couette flow, we set a constant streamwise velocity $U_w$ at the top wall, and use Dirichlet boundary conditions for all other velocity components at the walls. 
The temperature boundary conditions are $\Theta_C$ and $\Theta_H$ at the cold and hot walls, respectively. 
The time integration is performed using a 2nd-order semi-implicit BDF scheme (SBDF2 in Dedalus). 
A `+' superscript will indicate the normalization of the spatial variables by the viscous length scale $\mu_{ref}^*/(\rho_{ref}^* u_\tau^*)$. 
The statistics reported in figure \ref{fig:MinChan} are obtained by averaging across $T^+=500$ time units in intervals of $dt^+=0.5$, obtained after transients. 

\begin{figure}
\subfigure{
\centering
\includegraphics[width=0.9\textwidth]{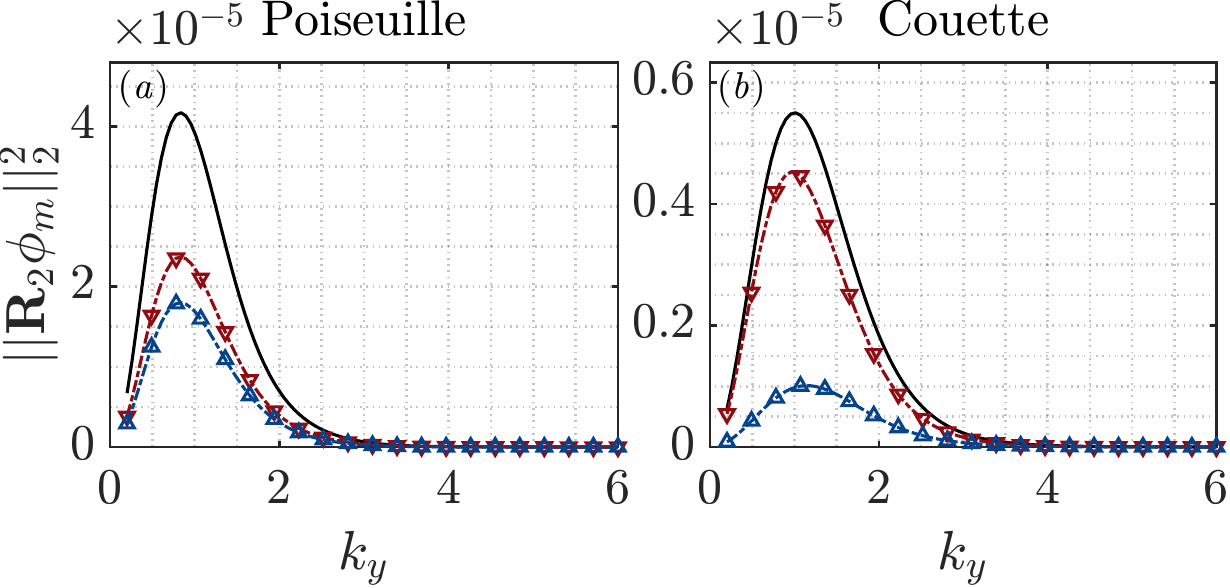}
\label{fig:Amp_R1R2_chan}
}%
\subfigure{
\centering
\label{fig:Amp_R1R2_couette}
}
\caption{
For (\textit{a}) Poiseuille and (\textit{b}) Couette flows, at $Pr=3$, $Re=500$ and $\Delta\Theta^* = 40K$, $||\bm{R}_2\bm{\phi}_m||_2^2$ is shown as a function of $k_y$. 
The responses across the full channel (solid lines) and across channel halves (dashed lines) are shown. 
The more-viscous ($\triangle$ markers) and less-viscous ($\nabla$ markers) channel halves are shown.  }
\label{fig:Amp_R2}
\end{figure}
The Poiseuille flow is considered at a fixed Reynolds number of $Re_\tau=186$. 
The streamwise, spanwise and wall-normal domain sizes are $L_x^+=350$, $L_y^+=350$ and $L_z^+=372$, respectively, with $N_x=32$, $N_y=64$ and $N_z=80$ grid points in each direction.
This gives grid spacings of $\Delta x^+=10.9$, $\Delta y^+=5.4$ and $0.14 \leq \Delta z^+ \leq 7.3$ \citep{bae2021nonlinear}. 
For the Couette flow, we take $Re_\tau=100$. 
The lower $Re_\tau$ for the Couette flow (in comparison to the Poiseuille flow) does not impact the discussions here, but makes the simulation, with the recommended grid spacings in the literature \citep{hamilton1995regeneration}, more computationally manageable. 
The streamwise, spanwise and wall-normal domain sizes are $L_x^+=549$, $L_y^+=376$ and $L_z^+=200$, respectively, and we use $N_x=48$, $N_y=48$ and $N_z=80$ grid points in each direction.  
This gives grid spacings of $\Delta x^+=11.4$, $\Delta y^+=7.8$ and $0.077 \leq \Delta z^+ \leq 3.92$. 

It should be emphasized that minimal flow unit simulations do not represent flows that can be realized in a laboratory. 
They instead serve as computationally tractable models that can be used to study some features of turbulent flows \citep{jimenez1991minimal}. 

\textbf{\textit{Route R2:}} Figure \ref{fig:Amp_R2} shows the amplification of $\bm{R}_2$ as a function of $k_y$. 
We see that $\bm{R}_2$ for the Poiseuille flow in figure \ref{fig:Amp_R2}(a) shows higher amplification at the less-viscous wall, and therefore does not exhibit the counterintuitive trend of higher viscosity enhancing turbulence.  

\textbf{\textit{Comparison with Sutherland's law of viscosity:}}

\begin{figure}
\captionsetup[subfigure]{labelformat=empty,skip=0pt}
\subfigure{
\centering
\centering
\includegraphics[width=0.9\textwidth]{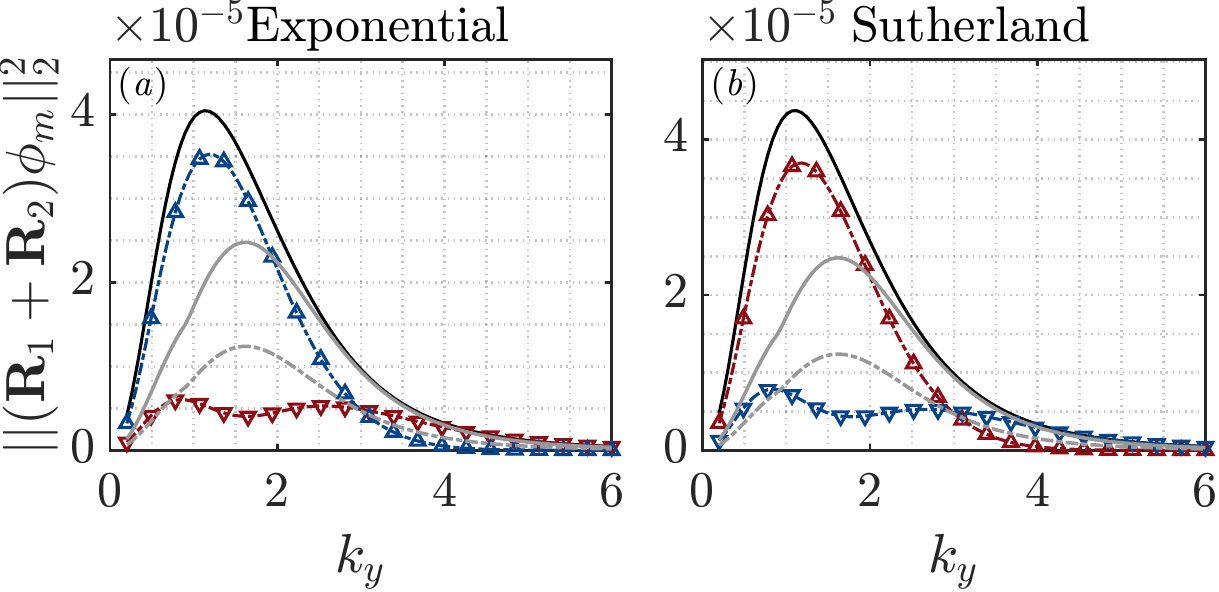}
\label{fig:Amp_R1R2_chan20}
}%
\subfigure{
\centering
\label{fig:Amp_R1R2_chanSuth}
}
\caption{
The same figure as in \ref{fig:Amp_R1R2} for Poiseuille flows at $Pr=3$ and $Re=500$ with (\textit{a})  exponentially varying viscosity with $\Delta\Theta^*=20K$ (\textit{b}) Sutherland's law based viscosity with $\Delta\Theta^*=160K$.  }
\label{fig:Amp_R1R2_suth}
\end{figure}

The numerical results in this study have been obtaining using the exponential variation of viscosity with temperature \eqref{eqn:visc} that is an empirical relation frequently employed for liquids like water. 
Figures \ref{fig:Amp_R1R2_suth} and \ref{fig:MinChan_ExpVSSuth} show that the conclusions drawn are also valid when considering the Sutherland's law of viscosity:
\begin{equation}
    \mu(\theta) = \theta^{3/2} \frac{(1+C)}{(\theta+C)},
\end{equation}
where $C = 110.4K/\Theta^*_{r}$. 
The key difference between the two functions for viscosity is that, while with the exponential law, the cold channel half is more-viscous than the hot half, for the Sutherland's law the opposite trend holds, with the cold half being less-viscous than the hot half. 
Here we choose $\Delta\Theta^*=20K$ for the exponential law and $\Delta\Theta^*=160K$ for the Sutherland's law, such that the ratio of the minimum to the maximum viscosity remains similar between the two cases. 

Figures \ref{fig:Amp_R1R2_suth} shows $||\bm{R}_1+\bm{R}_2||_\infty$ for the Poiseuille flow at $Pr=3$ for the exponential law and the Sutherland's law. 
We see that the trends are strikingly similar with the more-viscous channel half showing higher streak amplification. 
Now considering the streamwise variance obtained from minimal channel simulations in figure \ref{fig:MinChan_ExpVSSuth}, we again see that the trends are similar, with the more-viscous half showing higher variance. 
This illustrates that the discussions here remain valid for the exponential law used for liquids, and the Sutherland's law used for gases. 

\begin{figure}
\captionsetup[subfigure]{labelformat=empty,skip=-0pt}
\subfigure{
\centering
\centering
\includegraphics[width=0.9\textwidth]{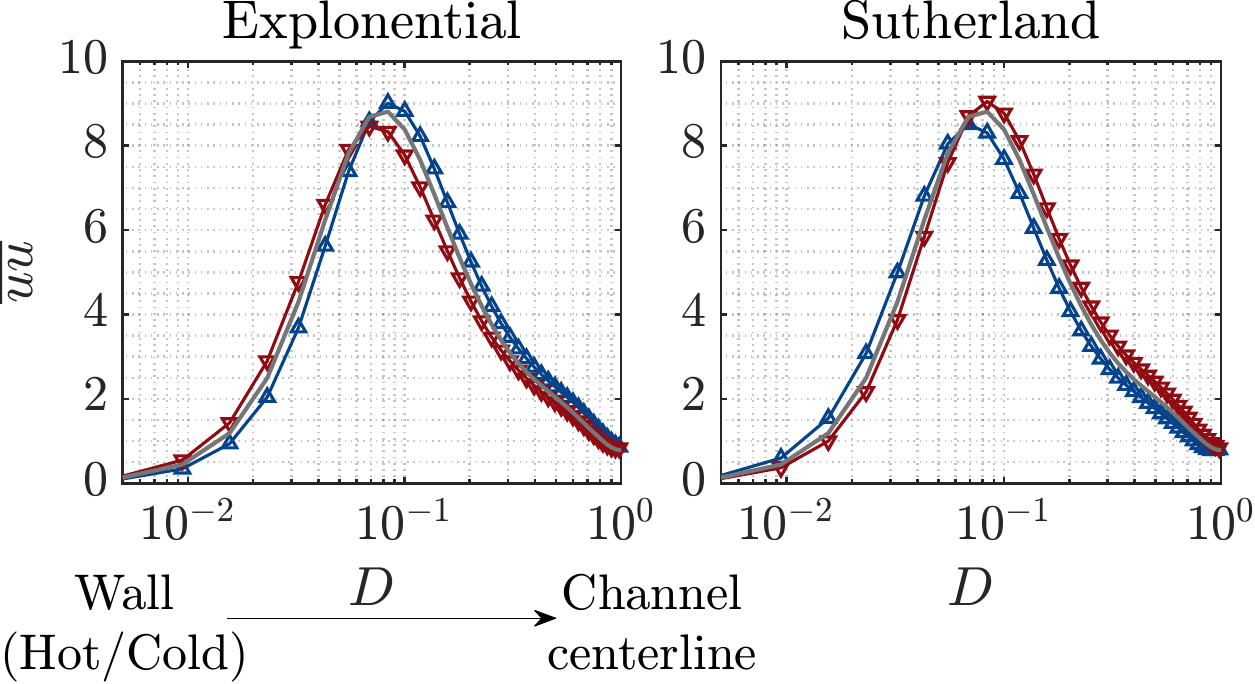}
\label{fig:MinChan_chan20}
}%
\subfigure{
\centering
\label{fig:MinChan_suthchan160}
}
\caption{
The same figure as \ref{fig:MinChan} obtained for Poiseuille flows at $Re_\tau=180$ and $Pr=1$ with (\textit{a})  exponentially varying viscosity with $\Delta\Theta^*=20K$ (\textit{b}) Sutherland's law based viscosity with $\Delta\Theta^*=160K$. 
}
\label{fig:MinChan_ExpVSSuth}
\end{figure}

\end{twocolumngrid}

\end{document}